\documentclass[aps,twocolumn,floatfix,showpacs]{revtex4}
\usepackage{graphicx,bm}

\begin{document}
\title{Bose-Einstein condensates with a bent vortex in rotating traps}
\author{Michele Modugno}
\affiliation{ 
 INFM - LENS - Dipartimento di Fisica, Universit{\`a} di Firenze,\\
 Via Nello Carrara 1, 50019 Sesto Fiorentino, Italy.}
\author{Ludovic Pricoupenko}
\affiliation{Laboratoire de Physique Th\'{e}orique des Liquides,
Universit\'{e} Pierre et Marie Curie,
case 121, 4 place Jussieu, 75252 Paris Cedex 05, France.}
\author{Yvan Castin}
\affiliation{Laboratoire Kastler Brossel, \'Ecole normale 
sup\'erieure, 24 rue Lhomond,
75005 Paris, France.}

\date{\today}

\begin{abstract}
We consider a 3D dilute Bose-Einstein condensate at thermal equilibrium in a rotating
harmonic trap. The condensate wavefunction is a local minimum of the Gross-Pitaevskii
energy functional and we determine it numerically with the very efficient conjugate
gradient method. For single vortex configurations 
in a cigar-shaped harmonic trap
we find that the vortex line
is bent, in agreement with the numerical prediction
of Garcia-Ripoll and Perez-Garcia, Phys.\ Rev.\ A {\bf 63}, 041603 (2001). 
We derive a simple energy functional for the vortex line in a cigar-shaped
condensate which allows to understand physically why the vortex line bends
and to predict analytically the minimal rotation frequency required
to stabilize the bent vortex line. 
This analytical prediction is in excellent agreement
with the numerical results. It also allows to find in a simple
way a saddle point of the energy, where the vortex line is in a stationary
configuration in the rotating frame but not a local minimum of energy.
Finally we investigate numerically
the effect of thermal fluctuations on the vortex line for a condensate
with a straight vortex: 
we can predict what happens in a single realization of
the experiment by a Monte Carlo 
sampling of an atomic field quasi-distribution function
of the density operator of the gas at thermal equilibrium in the Bogoliubov
approximation.
\end{abstract}


\maketitle

\section{Introduction}
Several experimental groups have now produced vortices in Bose condensates of
atomic gases by a rotation of the trapping potential \cite{ens,MIT,Foot}.
It is therefore important to characterize the steady state vortex configurations
for parameters relevant to the experiments. In particular the regime of the ENS and MIT
still deserves some investigation. It corresponds to a condensate
trapped in a cigar-shaped harmonic potential, that is with a harmonic oscillation
frequency along the rotation axis $z$ smaller by one order of magnitude than
the (quasi-degenerate) oscillation frequencies in the $x-y$ plane.
Furthermore, a configuration with a single vortex can be produced in the experiment
in a reproducible way \cite{ens}. The single vortex configuration
in a cigar-shaped trap is the main object of the present work.

Why is the single vortex cigar-shaped
regime so interesting? First the weak harmonic confinement along $z$ 
makes the system very different from previously studied configurations.
If the harmonic confinement along $z$ was stronger than in the $x-y$ plane
the vortex configurations would be similar to what happens in a 2D rotating Bose
gas, a well studied limiting case \cite{Fetter_review,Rokhsar,castindum,du}. 
If the harmonic confinement along $z$
was absent, {\it e.g.} if the condensate was a cylinder, one would face
a situation typical of superfluid helium II, a well studied subject \cite{Donnelly}.

But now the gas keeps its 3D character while being spatially inhomogeneous
along $z$. This has the amusing consequence that a single vortex line has a
tendency to bend. First the steady state condensate can have a bent vortex
line, as shown numerically by a minimization of the Gross-Pitaevskii
energy functional in \cite{garcia} and as obtained also numerically
with an approximate vortex line energy functional in \cite{AftaRiv}. 
Second, even a condensate with
a straight vortex line has low energy Bogoliubov modes corresponding to large
fluctuations of the end points of the vortex line \cite{Fetter_dynamics,feder}
so that the vortex line can easily bend in presence of thermal fluctuations.

After a presentation of the model considered in this article, see section
\ref{sec.model}, we give
a systematic numerical study of the steady state bending as function of the trap
aspect ratio, see section \ref{sec:numerics}, 
an analytic understanding of the bending of the steady state vortex line
in a cigar-shaped trap, see section \ref{sec:analytics}, 
and the description of the finite temperature fluctuations of the vortex line,
see section \ref{sec:finite_temp},
including a discussion of the effect of vortex line bending on the experimental
absorption images.

\section{Model and basic assumptions}
\label{sec.model}
Let us consider an almost pure Bose-Einstein condensate of $N$ atoms confined in a 
trapping potential rotating along axis $z$ at angular 
velocity $\Omega$.  The thermodynamically metastable equilibrium
configurations of the
system correspond to local minima of the Gross-Pitaevskii 
energy functional 
in the rotating frame \cite{castindum}
\begin{equation}
\label{eq:ener}
E[\phi,\phi^*] = \int d^3\vec{r}\,\left[
\phi^*\left(H_0-\Omega L_z\right)\phi
+\frac{N g}{2} |\phi|^4 \right],
\end{equation}
where the condensate wave function $\phi$ obeys the normalization condition
\begin{equation}
\label{eq:norm}
||\phi||^2\equiv \langle\phi|\phi\rangle=\int d^3\vec{r}\, |\phi|^2 =1.
\end{equation}
The Hamiltonian operator $H_0$ in Eq. (\ref{eq:ener})
contains the kinetic and trapping potential terms
\begin{equation}
\label{eq:h0}
H_0 = -\frac{\hbar^2\nabla^2}{2m} + U(\vec{r}\,)
\end{equation}
and $L_z=-i\hbar(x\partial_y-y\partial_x)$ is the angular momentum
operator along $z$. 
Here we will consider harmonic trapping potentials $U(\vec{r}\,)$ 
with an adjustable slight anisotropic deformation in the $x-y$ plane:
\begin{equation}
U(\vec{r}\,) = \frac{1}{2} m\omega_\perp^2\left[(1-\epsilon)x^2
+(1+\epsilon)y^2\right] + \frac{1}{2} m\omega_z^2 z^2.
\end{equation}
For the choice of parameters we refer to the typical values
of the recent ENS experiments with $^{87}$Rb atoms
(scattering length $a=100\,a_0\simeq5.29\times 10^{-9}m$) \cite{ens}: 
total number of atoms $N=1.4\times 10^5$, 
axial frequency $\omega_z=2\pi\times11.7\,{\rm Hz}$ and 
anisotropy parameter $\epsilon\ll1$. 
In this paper the radial frequency $\omega_\perp$
is varied in order to explore a wide range of trap anisotropies:
{$\lambda^{-1}\equiv\omega_\perp/\omega_z \in [1,15]$}.
In what follows energies and lengths are given in units of
$\hbar\omega_\perp$ and $a_\perp=\sqrt{\hbar/m\omega_\perp}$ respectively,
$m$ being the atomic mass of $^{87}$Rb.

\section{Local minima of energy: numerical results}
\label{sec:numerics}

In this section we discuss the method and the algorithm 
employed to minimize numerically the energy functional (\ref{eq:ener}).
Then we present the results obtained for the stationary configurations
with and without vortices in a long cigar trap, and finally we discuss
the existence domain for a single vortex configuration for several
trap geometries.

\subsection{Method}
First of all, we reformulate the energy functional in order to account
more easily for
the normalization constraint (\ref{eq:norm}): we define 
$\phi=\psi/||\psi||$ so that the value of $E$ can be obtained
for condensate wave functions $\psi$ with a norm
different from unity. This corresponds to dividing the terms
of $E[\psi,\psi^*]$ quadratic in $\psi$ by $||\psi||^2$,
and the interaction term quartic in $\psi$ by $||\psi||^4$.
The modified energy functional reads
\begin{equation}
\label{eq:enerpsi}
E[\psi,\psi^*] = \int d^3\vec{r}\,
\frac{\psi^*\left[H_0-\Omega L_z\right]\psi}{||\psi||^2}
+\frac{N g}{2} \frac{|\psi|^4}{||\psi||^4}. 
\end{equation}
Then we discretize $\psi$ on a three-dimensional grid 
with periodic boundary conditions in position and in momentum space. 
The number of grid points that we use ranges from 64 to 256
for each lattice side, depending on the trap geometry (a typical choice is 
$64\times64\times192$ for long cigar configurations and 
$96\times96\times64$ for spherical geometries with vortices).
The minimization is performed by using the {\em conjugate gradient} 
algorithm described in \cite{numrec}. 
This algorithm is in general much more efficient
that the usual steepest descent method (see Ref. \cite{numrec} for 
a comprehensive discussion).
One ingredient of the conjugate gradient method is a line minimization
of the energy functional, that is the minimization of $E$ along
the line $\psi = \psi_0 + \lambda \delta\psi$, $\lambda$ being a real
parameter, where $\psi_0$
is the current trial wave function and  $\delta\psi$ is the proposed direction
(or gradient) along which to move the trial wave function.
An important issue here is to find the {\it first} 
minimum encountered when moving downhill in energy
along the line: if the algorithm can jump to another minimum on the line,
corresponding for example 
to an energy valley with a different number of vortices,
one gets wrong predictions, in the sense that one
does not fully explore the stability domain of a branch of solution
with a given number of vortices. This issue is usually not considered
as important in the textbook implementation of the conjugate gradient
method: {\it e.g.}, in \cite{numrec} the line minimization does not
necessarily find the first minimum. 
What we have done is to use the fact that $E$ is a rational
function of $\lambda$ so that it is completely characterized 
by the coefficients of the monomials in $\lambda$ 
in the numerator and the denominator.
We then easily find the roots of
$dE/d\lambda$ and the first local minimum of $E$ encountered when one moves
along the line downhill in $E$ starting from $\lambda=0$. Afterwards we proceed
with another line minimization along a {\em conjugate direction} \cite{numrec},
and so forth until we find a local minimum of the energy functional 
(\ref{eq:enerpsi}) in the full configuration space spanned
by the wave function $\psi$.
As convergence criterion we use 
\begin{equation}
||H_{\rm gp} \psi -\mu \psi|| < \gamma \mu ||\psi||
\label{eq:criterion}
\end{equation}
where $H_{gp}$ is the Gross-Pitaevskii Hamiltonian
\begin{equation}
H_{\rm gp}\equiv H_0 + Ng |\psi|^2/||\psi||^2 
\end{equation}
and $\mu$ is given by
\begin{equation}
\mu = \int d^3\vec{r} \frac{\psi^* H_{\rm gp} \psi}{||\psi||^2}
\end{equation}
which eventually converges to the chemical potential. The parameter $\gamma$ 
is a small parameter ($\sim10^{-8}\div10^{-10}$) controlling the quality 
of the convergence.

\subsection{Results for a fixed trap geometry: case of a long cigar}
\label{sec:longcigar}

\begin{figure}
\centerline{\includegraphics[width=8cm,clip=]{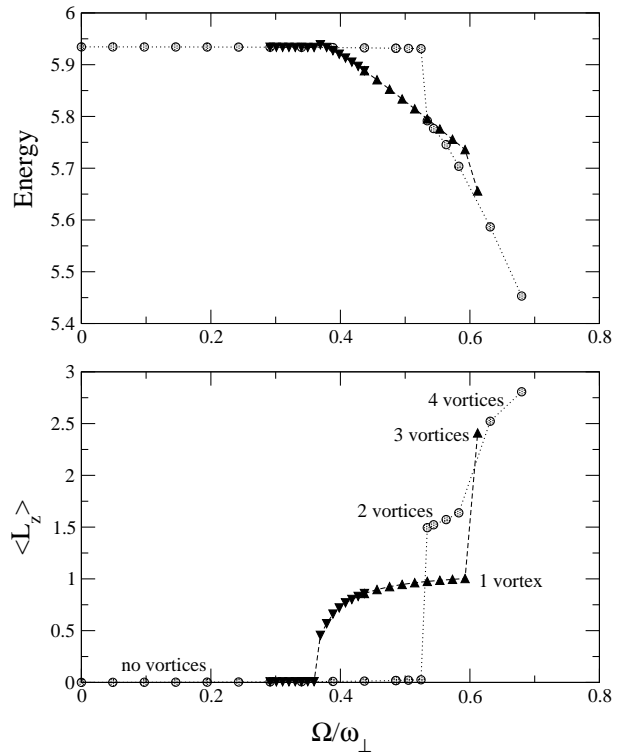}} 
\caption{Energy per particle (top) and angular momentum per particle 
$\langle L_z \rangle $ (bottom)
as a function of the rotation frequency $\Omega$
for various branches. Circles: starting from the $\Omega=0$ 
ground state without vortices
we reach configurations with 2 and 4 vortices as we increase $\Omega$. 
Triangles: configurations with one bent vortex which decays into a
3 vortex (no vortex) state as we increase (decrease) $\Omega$. This branch
is obtained by starting from a trial wave function with quantized
circulation of charge $+1$ for a value of $\Omega$ which can 
support a vortex state. See the text for the value of the parameters.
The energy is in units of $\hbar\omega_\perp$ and $\langle L_z\rangle$
in units of $\hbar$.} 
\label{fig:lz}
\end{figure}

We start by considering a cigar shaped trap geometry typical of the
ENS experiments in which vortex configurations are nucleated
\cite{ens,feder}. 
In particular we set $\epsilon=0.03$ and
$\omega_{\perp}=2\pi\times174$Hz, giving
$\lambda^{-1}\simeq14.9$.
To explore the configuration space we start from a trial wave function
for a fixed value of the rotation frequency $\Omega$ until an equilibrium
configuration ({\em i.e.} a local minimum of the energy functional
(\ref{eq:enerpsi})) is found. 
Then we increase or decrease slightly 
$\Omega$, and then find the new local minimum. In this way we can
follow continuously branches of
configurations with or without vortices, depending on the rotation
frequency and on the path followed. 
In Figure \ref{fig:lz} we show the energy $E$ and the angular momentum per particle
$\langle L_z \rangle$ along the rotation axis of each configuration, as function 
of the rotation frequency $\Omega$, up to 4-vortex configurations.  

\begin{figure} 
\centerline{\includegraphics[width=5cm,clip=,angle=-90,
bb=160 210 465 625]{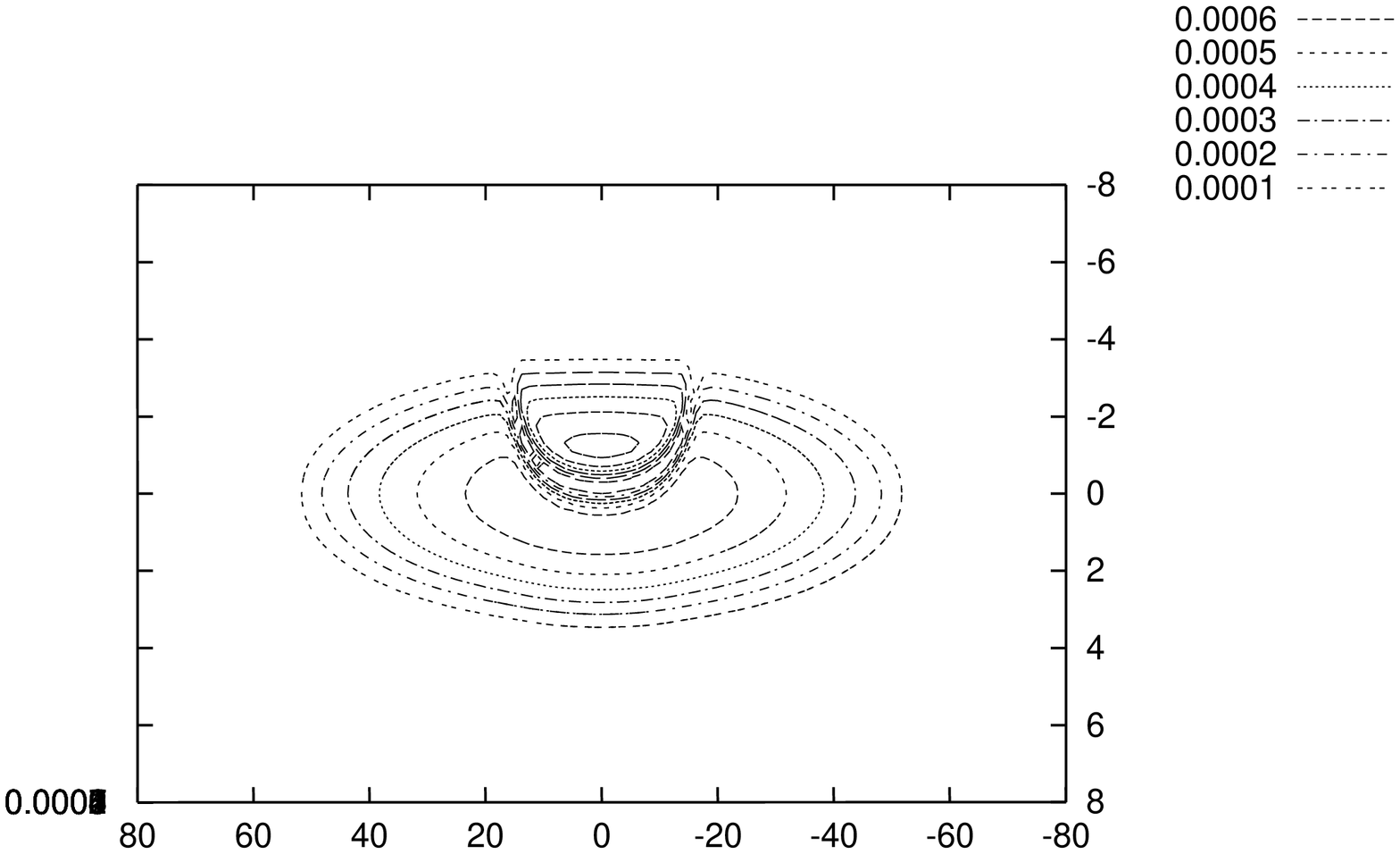}}
\centerline{\includegraphics[width=5cm,clip=,angle=-90,
bb=160 210 465 625]{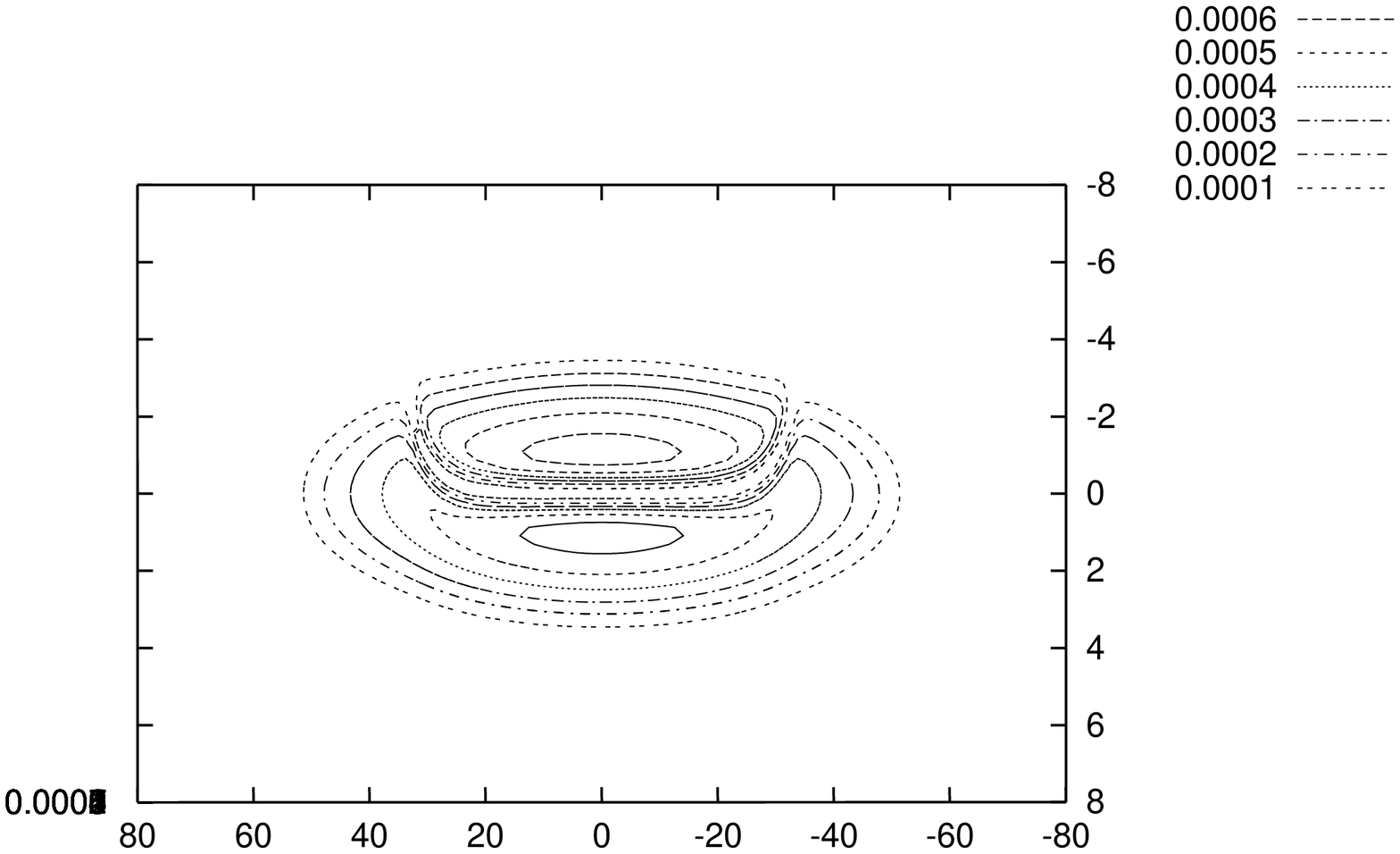}}
\centerline{\includegraphics[width=5cm,clip=,angle=-90,
bb=160 210 465 625]{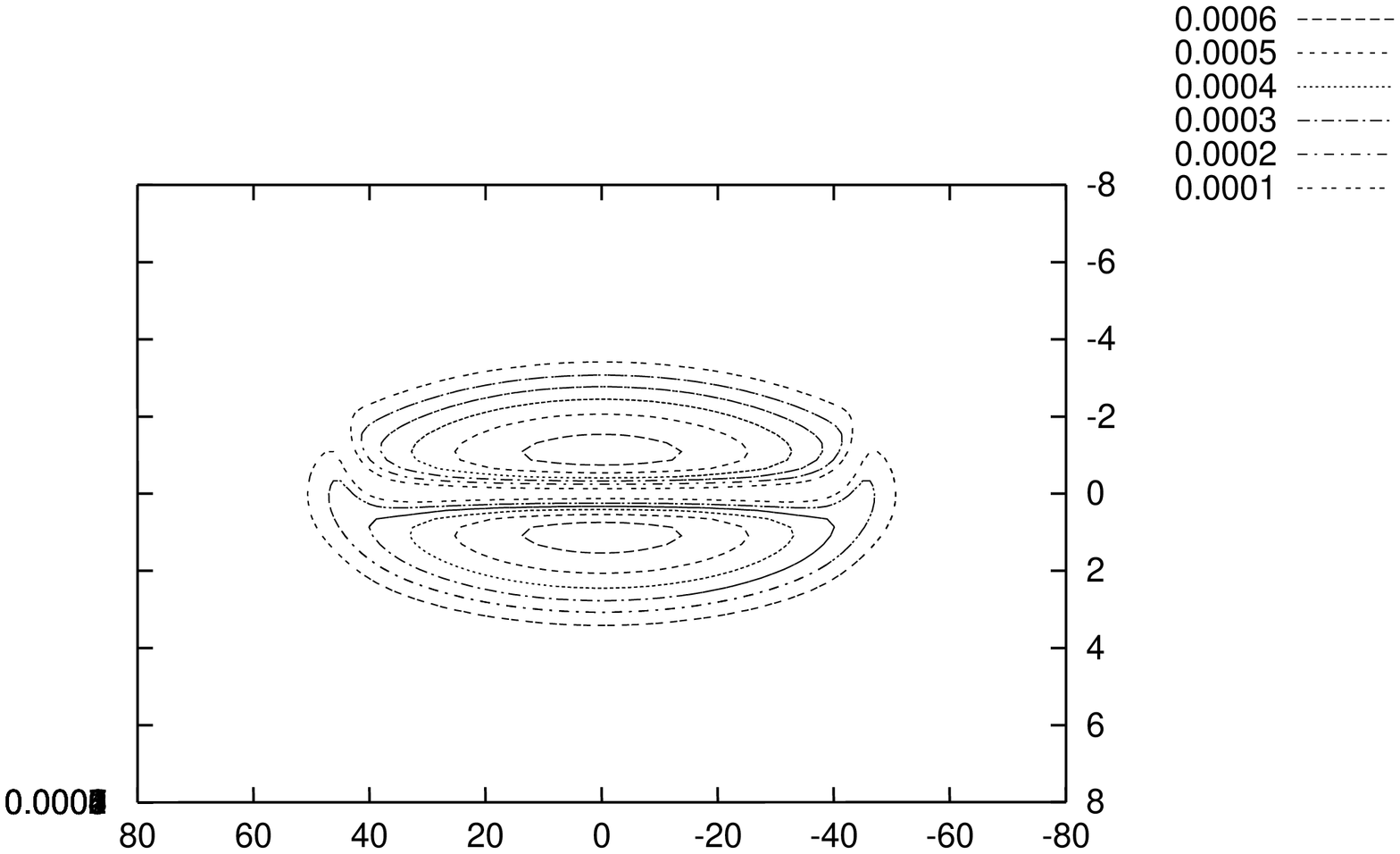}}
\caption{Density cuts in $y-z$ plane 
(vertical and horizontal axes respectively)
of a long cigar condensate ($\lambda^{-1}=14.9$) with a bent vortex, 
 for three values of the rotation frequencies: 
$\Omega/\omega_\perp=0.37, 0.44, 0.59$ from top to bottom.
Lengths are given in unit of $a_\perp$. }
\label{fig:3bent}
\end{figure}

We notice that, for this trap geometry, when a single vortex appears
({\em i.e.} when it becomes a stable configuration) it
has immediately a lower energy than the 0-vortex configuration, 
differently to what happens in 2D \cite{castindum}. 
Moreover this vortex configuration
is characterized by a bending of the core line, a phenomenon already
obtained with a different numerical technique in \cite{garcia},
and this accounts for the fact that the angular momentum per particle
$\langle L_z \rangle$ is lower
than $\hbar$ ($\langle L_z \rangle $ is equal to $\hbar$ 
for a straight centered vortex).
To better investigate this aspect we have studied the deformation of 
the vortex line as function of the rotating frequency. 
If we start with a single vortex configuration and we
decrease the rotation frequency $\Omega$ the bending of the
core line becomes more pronounced, until we reach a certain
critical frequency  $\Omega_1$ and the system jumps
to the 0-vortex configuration. In the opposite direction,
when $\Omega$ increases, we find that the vortex line
tends to become more straight as expected (see Figure \ref{fig:3bent}),
but then, whereas a small bending is still present at the extremities of 
the condensate, it ``decays'' into a three vortex configuration.
The value of $\Omega$ at which this happens  defines a second critical 
rotation frequency  $\Omega_2$.

Therefore for the long cigar trap and the parameters considered
here, one never obtains a single, straight vortex: 
when $\Omega$ is too large, other vortices come in. 
As we will show in the following, this is due
to destabilization of high angular momentum surface modes \cite{dalfovo},
and corresponds to the fact that the single vortex configuration
ceases to be a local minimum of energy when the energy of these modes 
becomes negative
({\em i.e.} there is at least one direction along which the energy
is going down in the functional space). 
We discuss this effect in details in the next subsection.

\subsection{Destabilization of surface modes}
\label{subsec:sumrules}

We consider in this subsection the case of a cylindrically symmetric trapping
potential (with a vanishing asymmetry parameter $\epsilon=0$)
so that eigenstates of the $N$-body Hamiltonian have a well
defined angular momentum.
The expression for the energy ${\cal E}_l$ of an
excitation mode of the condensate with angular momentum $\hbar l$
in the rotating frame is
\begin{equation}
{\cal E}_l=\hbar\omega(l)-l\hbar\Omega,
\label{eq:E_omega}
\end{equation}
where $\omega(l)$ is the excitation energy of the same mode in 
the laboratory frame \cite{castindum}. This determines the rotation 
frequency $\tilde{\Omega}_l\equiv\omega(l)/l$
 at which the energy of such mode becomes 
negative in the rotating frame. 
In particular  the minimum value over $l$ of $\tilde{\Omega}_l$ defines the
Landau critical rotation frequency $\tilde{\Omega}$ at which the 
vortex becomes thermodynamically unstable due to destabilization 
of these surface modes
\begin{equation}
\tilde{\Omega}=\min_{\{l\}}\left(\frac{\omega(l)}{l}\right).
\end{equation}
The value $\tilde{\Omega}$ can be estimated analytically by using 
the sum rules approach which provides an estimate for $\omega(l)$,
as discussed in Refs. \cite{dalfovo,zambelli}. In order
to describe the surface modes, we consider the excitation operators
for $l\neq0$
\begin{equation}
F_{\pm} = \sum_{k=1}^{N} (x_k\pm i y_k)^l.
\end{equation}
Then one defines the moments $m_p^\pm$
\begin{equation}
m_p^\pm\equiv\int_0^{+\infty}dE\left[S_+(E)\pm S_-(E)\right]E^p
\label{eq:mp}
\end{equation}
where the strength distribution functions $S_\pm$ are given by
\begin{equation}
S_\pm(E)\equiv\sum_n|\langle n|F_\pm|0\rangle|^2\delta(E-\hbar\omega_n).
\label{eq:strenght}
\end{equation}
The states $|n\rangle$ form a complete set of eigenstates of the
Hamiltonian operator $H$ for our system of $N$ interacting particles
confined by the trapping potential:
\begin{equation}
\label{eq:hamilt}
H = \sum_{k=1}^{N}\left[\frac{p_k^2}{2m} + U(\vec{r}_k\,) +g\sum_{j=1}^{k-1}
\delta(\vec{r}_j-\vec{r}_k)\right].
\end{equation}
Here we assume the two mode approximation, 
as discussed in Ref. \cite{zambelli}:
\begin{equation}
S_\pm(E)=\sigma\delta(E-\hbar\omega_\pm),
\end{equation}
where the two modes are equally weighted (they have the same 
strength $\sigma$) due to the vanishing of the $m_0^-$ 
momentum \cite{zambelli}. This is justified by the fact
 that in the large $N$ limit the strength distributions
are dominated by the contribution of two modes with energy
$\hbar\omega_\pm$ and angular momentum $\pm\hbar l$ excited respectively 
by the operators $F_\pm$.
With this ansatz it is straightforward to obtain the relation between
the frequencies $\omega_\pm$ of such modes and the first three moments
in Eq. (\ref{eq:mp}); in particular we have
\begin{eqnarray}
\delta&\equiv&\omega_+-\omega_-=\frac{m_2^-}{m_1^+}
\label{eq:delta} \\
\bar{\omega}^2&\equiv&\left(\frac{\omega_++\omega_-}{2}\right)^2=
\frac{m_3^+}{m_1^+}-\frac{3}{4}\delta^2.
\label{eq:wbar}
\end{eqnarray}
Notice that only the mode with frequency $\omega_+=\bar{\omega}+\delta$
is relevant to our discussion, since the energy of the mode with 
angular momentum $-\hbar l$ increases for increasing rotation frequency 
$\Omega$, and can never become negative.

The next step is to calculate the moments $m_p^\pm$ from Eqs. (\ref{eq:mp})
and (\ref{eq:strenght}) by using closure relations and then explicitly 
evaluating the commutators
\begin{widetext}
\begin{eqnarray}
m_1^+&=&\langle[F_-,[H,F_+]]\rangle
=\frac{2N\hbar^2}{m}l^2\langle r_\perp^{2l-2}\rangle
\\
m_2^-&=&\langle[[F_-,H],[H,F_+]]\rangle
=\frac{4N\hbar^3}{m^2}l^2(l-1)\langle r_\perp^{2l-4}L_z\rangle
\\
m_3^+&=&\langle[F_-,H],[H,[H,F_+]]\rangle
=\frac{2N\hbar^4\omega_\perp^2}{m}l^3\cdot\nonumber
\\
&&\cdot\left[
\langle r_\perp^{2l-2}\rangle+\frac{l-1}{m^2\omega_\perp^2}\cdot
\frac{\langle p_+r_\perp^{2l-2}p_-\rangle + 
2\hbar(l-2)\langle r_\perp^{2l-6}L_z\rangle+ 
2(l-2)\langle r_\perp^{2l-6}L_z^2\rangle/l}
{\langle r_\perp^{2l-2}\rangle}\right]
\end{eqnarray}
where $\langle\ldots\rangle$ stands for the expectation value
in the state $|0\rangle$.
These expressions generalize those in \cite{dalfovo}
(case of a vortex free condensate) and in \cite{zambelli}
(case of excitations of $l=2$ of a condensate with vortices).
We note that the use of $p_\pm\equiv p_x\pm i p_y$ and $r_\pm\equiv
x\pm i y$ rather than the Cartesian coordinates $p_x$, $p_y$, $x$,
$y$, {\em e.g.} by writing the kinetic energy as $p_+p_-$,
considerably simplifies the calculations of the commutators.

For the case of a straight vortex with charge $q$ the key quantities
to be inserted in Eqs. (\ref{eq:delta}) and (\ref{eq:wbar}) are
\begin{eqnarray}
\frac{m_2^-}{m_1^+}
&=&2q(l-1)\omega_\perp\left[\frac{\langle r_\perp^{2l-4}\rangle}
{\langle r_\perp^{2l-2}\rangle}\right]
\label{eq:sumr1}
\\
\frac{m_3^+}{m_1^+}
&=&\omega_\perp^2 l\left[1+(l-1)\frac{\langle p_+r_\perp^{2l-4}p_-\rangle}
{\langle r_\perp^{2l-2}\rangle}+\frac{2q(l+q)(l-2)(l-1)}{l}
\frac{\langle r_\perp^{2l-6}\rangle}
{\langle r_\perp^{2l-2}\rangle}\right],
\label{eq:sumr2}
\end{eqnarray}
\end{widetext}
where quantities within square brackets are expressed in dimensionless units, 
as defined in Sect. \ref{sec.model}.

\begin{figure} 
\centerline{\includegraphics[width=8cm,clip=]{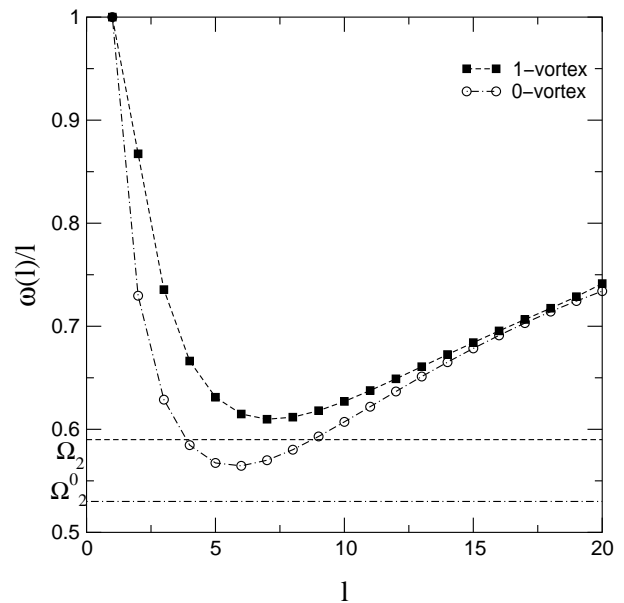}}
\caption{Curve $\omega(l)/l$ for a condensate in a long cigar trap
($\lambda^{-1}=14.4$, $\epsilon=0$) 
with (squares) and without (circles) a straight vortex, 
as calculated with the sum rules approach 
(see Eq. (\protect\ref{eq:sumr1}) and (\protect\ref{eq:sumr2})).
The minimum of $\omega(l)/l$, which gives an estimate for 
the rotation frequency $\Omega_2$, is compared with its corresponding 
value found numerically for the 0-vortex ($\Omega^0_2$, dot-dashed line) and 
the 1-vortex ($\Omega_2$, dashed line) states, in presence of a
slight anisotropy ($\epsilon=0.03$). Frequencies are given
in units of $\omega_\perp$.}
\label{fig:sumrules}
\end{figure}

In Figure \ref{fig:sumrules} we plot the curve $\omega(l)/l$ for a
cylindrically symmetric trap ($\epsilon=0$), for the case
$q=0$ (ground state without vortices) and $q=+1$ (straight
vortex). In the same figure we also indicate the critical frequencies
$\Omega^0_2$ and $\Omega_2$ which define the upper bound of the existence 
domain respectively for states without vortices and with a bent
vortex, as found from the
numerical minimization of the energy functional in presence of a
slight anisotropy as considered in the previous subsection ($\epsilon=0.03$). 
We notice that the sum rule prediction gives a value 
of $\min\left(\omega(l)/l\right)$ which is quite close to our 
numerical result.
This supports the statement that the surface modes are indeed responsible 
for the destabilization of the single-vortex state and the vortex-free state.

\subsection{Effect of a ballistic expansion}
\label{subsec:tof}

In general, in the experiments, the condensate density is imaged 
after a time-of-flight: the trapping potential is switched off
and the cloud expands freely for some time before being imaged
by the absorption of a laser beam. This has the advantage of
increasing the diameter of the vortex core, making it larger than
the optical wavelength. Some theoretical work is however needed
to relate the images obtained after the expansion to the density
profile of the initial trapped condensate.

To calculate the expansion of the condensate 
after the release from the trap,
we propagate the initial wavefunction according
to the full 3D Gross-Pitaevskii equation in real time after
having abruptly switched off the trapping potential,
and after having performed 
the scaling and gauge transform of \cite{Castin-Dum,Kagan}.
In terms of the rescaled spatial coordinates 
$\tilde{r}_i\equiv r_i/\lambda_i(t)$, the modified wave function
$\tilde{\psi}(\vec{\tilde{r}},t)$ satisfies the equation
\begin{eqnarray}
i\partial_t\tilde{\psi}(\vec{\tilde{r}},t)&=&
\left[-\frac{\hbar^2}{2m}\sum_j{1\over\lambda^2_j}\partial^2_{\tilde{r}_j}
\right.\\
&&\left.+\frac{1}{\prod_j\lambda_j}\left(U(\vec{\tilde{r}})+g|\tilde{\psi}|^2
-\mu\right)\right]\tilde{\psi}(\vec{\tilde{r}},t)\nonumber
\end{eqnarray}
and the scaling parameters $\lambda_j(t)$ are solutions of \cite{Castin-Dum}
\begin{eqnarray}
\ddot{\lambda}_j&=&\frac{\omega_j^2(0)}{\lambda_j\lambda_1\lambda_2\lambda_3}
\\\nonumber
\end{eqnarray}
with initial values equal to unity and with vanishing initial
derivatives.
The use of these scaling equations allows us to calculate numerically
the expansion of any vortex configuration
by using the same grid as the one 
used for the stationary trapped state as most
of the evolution of $\psi$ due to the ballistic expansion is absorbed
by the scaling and gauge transform. In a 2D model with an axially symmetric
trapping potential the net effect 
of the ballistic expansion on the
density corresponds exactly to a scaling transformation and a finite angle
rotation of the condensate eigenaxes \cite{castindum}. Here we
expect this property to remain approximately true as the condensate
is very elongated along $z$ and expands much more rapidly radially 
than axially.

\begin{figure}
\centerline{\includegraphics[width=8cm,clip=]{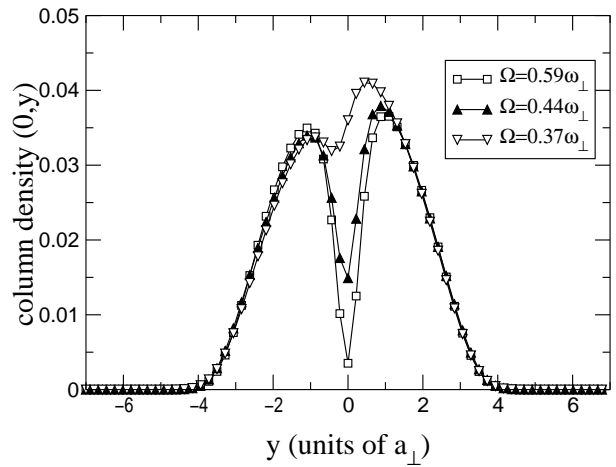}}
\caption{Column density before ballistic expansion
(integrated density along the axial direction)
 along the $y$ axis where the vortex bends,
for the three cases in Figure \protect\ref{fig:3bent}: 
$\Omega/\omega_\perp=0.37,0.44,0.59$. This picture evidences 
that the contribution of the vortex line bending to the contrast 
can be relevant even at zero temperature.}
\label{fig:profil_michele}
\end{figure}

\begin{figure}
\centerline{\includegraphics[width=8cm,clip=]{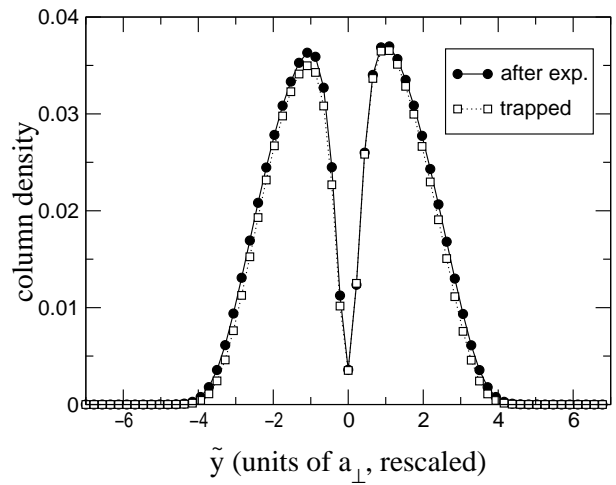}}
\caption{Comparison of the column density for a bent vortex 
($\Omega=0.59 \omega_\perp$) before and after the release from the trap
(time of flight $\sim30$ms), as a function of
the rescaled $\tilde{y}$ coordinate. The expansion has no relevant effects
on the contrast (we have verified that this holds also for
the other configurations shown in Figure \protect\ref{fig:3bent}).}
\label{fig:afterexp}
\end{figure}

A first application of this calculation
is the analysis of the column density of the condensate, that is
the density integrated along $z$, after ballistic expansion.
There is indeed an important issue concerning 
the mechanisms which are responsible for the
reduction of contrast for the vortex configurations which are observed
in the experiments \cite{ens}. As pointed out
already in \cite{garcia}, in the case of a bent vortex before
ballistic expansion,
an important contribution can arise due to the bending itself, 
as shown in Figure \ref{fig:profil_michele}. In this picture we plot a cut 
of the column density
along the $y$ axis where the vortex bends, for three values of the 
rotation frequency $\Omega$. 
In Figure \ref{fig:afterexp} we compare the column density for a bent vortex 
at $\Omega=0.59 \omega_\perp$, before and after the release from the trap
(time of flight $\sim30$ms). The $y$ coordinate is given in rescaled units
$\tilde{y}\equiv y/\lambda_2(t)$. From this figure we can see that the 
expansion has no relevant effects on the contrast; the same result 
holds also for the other configurations shown in Figure \ref{fig:3bent}.

\begin{figure}
\centerline{\includegraphics[width=8cm,clip=]{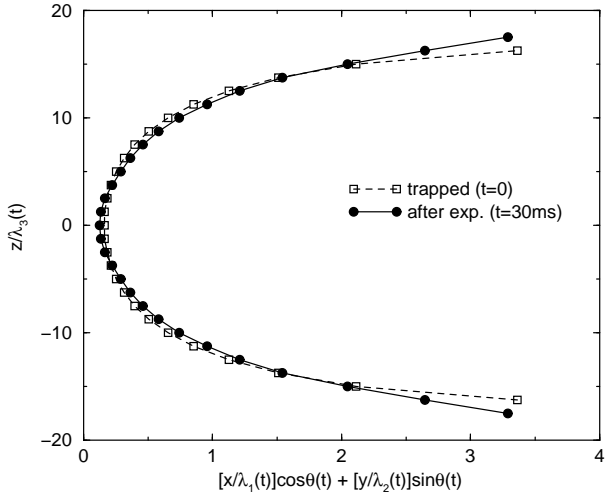}}
\caption{Shape of the vortex line (with initial
rotation frequency $\Omega=0.37 \omega_\perp$) 
before and after release from the trap
(time of flight $t_e\sim30$ms), as a function of
the rescaled coordinates in units of $a_\perp$
and after projection
on the vertical planes at an angle $\theta(0)=-\pi/2$ and 
$\theta(t_e)=-1.14$rad with $x$ axis, respectively.}
\label{fig:ligne}
\end{figure}

A second application is the analysis of the shape of the 
vortex line after ballistic expansion. This is also an important issue,
as the shape of the vortex line can be measured using transverse
absorption imaging rather than axial absorption imaging
\cite{vu_a_ENS}. Focusing
on the configuration $\Omega=0.37\omega_\perp$ exhibiting the 
strongest bending, we have calculated the shape of the vortex line
before and after expansion by looking in each horizontal
plane for the local minimum of density closest to the vertical axis $z$.
Initially the vortex line is contained in the half plane $(-y)z$ which is at
an angle $\theta(0)=-\pi/2$ with respect to $x$ axis.
After a time $t_e\simeq 30$ms of ballistic expansion, 
we find that the vortex line remains
almost planar, but in a vertical half plane at an angle $\theta(t_e)
\simeq -1.14$rad with respect to $x$ axis,
and not passing through the origin. We show in Figure \ref{fig:ligne}
the vortex line before and after expansion, projected in
the vertical planes in which it is contained at $t=0$ and
$t=t_e$.  One then sees that the vortex line essentially preserves its
initial shape after time of flight, in the frame of the
rescaled spatial coordinates.

\subsection{Existence domain of a single vortex configuration}

In the last part of this section we discuss the thermodynamic existence domain 
of a single vortex configuration, that is the domain where it is a
local minimum of energy. By varying the radial frequency
$\omega_\perp$ we have explored a wide region of trap geometries,
ranging from the spherical case to long cigar traps ($\lambda^{-1}
\in[1,15]$). 
\begin{figure} 
\centerline{\includegraphics[width=8cm,clip=]{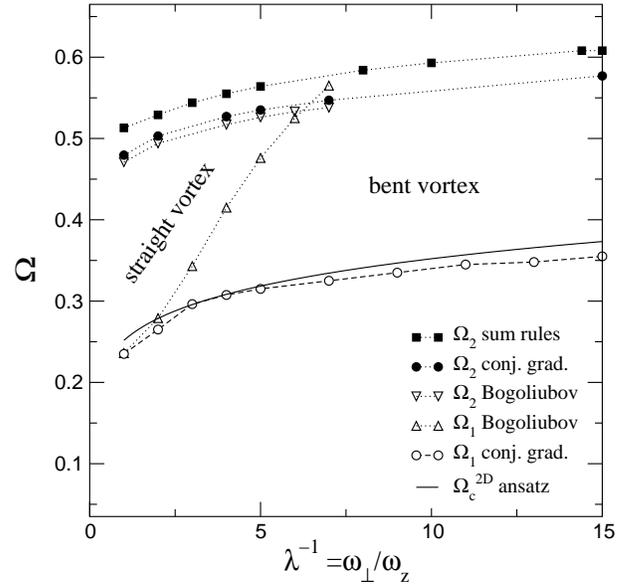}} 
\caption{Phase diagram for the existence domain of a single vortex
as a local minimum of energy in an almost cylindrically symmetric trap.
For a given $\lambda$, thermodynamically stable configurations 
lie in the interval $[\Omega_1,\Omega_2]$. 
The Bogoliubov predictions from $\Omega_1$ and $\Omega_2$ (triangles)
correspond to a straight vortex.
In the conjugate gradient minimization (disks)
the bending of the vortex line is allowed and the stability domain is enhanced.}
\label{fig:diagram}
\end{figure}
In Figure \ref{fig:diagram} we show this domain in the 
$\omega_\perp/\omega_z-\Omega/\omega_\perp$ plane. 
Here we consider an almost axially symmetric trap, with a very small
asymmetry ($\epsilon=10^{-4}$), which fixes the
direction along which the vortex bends.
By minimizing numerically the energy functional (\ref{eq:ener})
with the conjugate gradient method we find that thermodynamically 
stable configurations with one vortex lie in the interval between 
the lines with empty and full circles, representing respectively 
$\Omega_1$ and $\Omega_2$. The black squares represent the predictions
of the sum rules, which give a good estimate for  $\Omega_2$,
as discussed in {\S}\ref{subsec:sumrules}.
In the same picture we also show the results from the Bogoliubov approach
for a straight vortex, as described in {\S}\ref{sec:finite_temp}. 
Up-triangles correspond to the critical rotation frequency which stabilize
a straight vortex ($\Omega_1$); this line separates the existence domains
for straight and bent vortices. Down-triangles correspond
to the frequency at which the straight vortex is no longer
thermodynamically stable due
to the destabilization of the surface modes ($\Omega_2$). 
Notice that these values
are in very good agreement with the result of the conjugate gradient
in the whole range of existence of the straight vortex.

Finally, the solid line represent a 2D ansatz based on the analytical 
model in Ref. \cite{castindum}, as described in the next section. 
If we imagine our 3D condensate
as a collection of 2D slices, we can suppose that the 3D vortex 
is a stable configuration ({\em i.e.} the vortex line remains 
close to $z$ axis and does not get away) 
if the rotation frequency $\Omega$ at least exceeds the 2D critical rotation
frequency $\Omega_c^{2D}$ at which a 2D vortex in the central 
slice ($z=0$) becomes energetically favorable with 
respect to the solution without vortices, that is: 
$\Omega > \Omega_c^{2D}(z=0)$. The expression for $\Omega_c^{2D}$,
which will be derived in the next section, is
\begin{equation}
\tilde{\Omega}_c^{2D}=\frac{1}{\tilde{\mu}}\ln\left(C'\left(\tilde{\mu}+
\eta\right)\right)
\end{equation}
where the tilda indicates rescaling by $\omega_\perp$,
{\it e.g.} $\tilde{\mu}\equiv\mu/\hbar\omega_\perp$, $C'\simeq1.8011$, and
$\eta=1.9378$. Then, by
expressing $\tilde{\mu}$ in terms of $\lambda=\omega_z/\omega_\perp$
in the Thomas-Fermi approximation:
\begin{equation}
\tilde{\mu}=
\frac{1}{2}\left(15N\frac{a}{a_z}\right)^{2/5}\!\!\!\!\lambda^{1/5}\simeq
13.1\lambda^{1/5}
\label{eq:mu_explicit}
\end{equation}
we have
\begin{equation}
\tilde{\Omega}_c^{2D}=\frac{b}{\lambda^{1/5}}\ln\left(c\lambda^{1/5}+d\right)
\end{equation}
with $b\simeq0.0763$, $c\simeq23.6$ and  $d\simeq3.49$. 
As we can see from 
Figure \ref{fig:diagram} this simple ansatz gives a good
estimate of the critical frequency $\Omega_1$. 
This ansatz will be justified in the next section.

\section{Understanding the bending of the vortex line analytically}
\label{sec:analytics}

The previous numerical results show that the vortex line in a steady
state configuration is not necessarily straight when the condensate 
is cigar shaped along the rotation axis $z$, in accordance
with previous numerical results based on a different algorithm \cite{garcia}.
This however does not explain physically {\it why} the vortex bends.
To get the required physical understanding we derive an approximate energy
functional for the vortex line in the Thomas-Fermi limit in
the spirit of \cite{AftaRiv}, and we
minimize this energy functional with a simple variational ansatz.
We reach a very simple prediction for the minimal rotation frequency
required to stabilize a bent vortex, which is in good agreement with the full numerics 
when the condensate is cigar-shaped.

\subsection{Deriving a simple energy functional}
We restrict in what follows to the interesting regime
of a cigar shaped condensate, where $\omega_z\ll\omega_\perp$.
The first step is to transform 
the Gross-Pitaevskii energy functional (\ref{eq:ener}) into a
functional of the shape of the vortex line only. 
This assumes that
both the modulus and the phase of the condensate wavefunction $\phi$ 
can be expressed in terms of the shape of the vortex line only.
This is in general a formidable task, as the condensate density is not
uniform in a harmonic potential \cite{pas_uni}, but it is greatly simplified
if we restrict to the Thomas-Fermi limit $\mu\gg\hbar\omega_\perp$.
We present a rather detailed and pedestrian derivation in the appendix
\ref{appendix}, we give here only the main ideas.

In the Thomas-Fermi regime there is a clear
separation of spatial scales between the vortex core radius,
of the order of the healing length $\xi=(\hbar^2/m\mu)^{1/2}$,
and the transverse Thomas-Fermi radius of the condensate
$R_\perp = (2\mu/m\omega_\perp^2)^{1/2}$, where $\mu$ is
the chemical potential of the gas.
The total density can then be written as the product of a slowly varying
envelope function and of a narrow \lq hole' function defining the vortex
core \cite{Fetter_review,castindum}.  
We further assume that the rotation frequency $\Omega$ is of
the order of $\hbar\omega_\perp^2/\mu$.
As a consequence the rotational velocity term
$\vec{\Omega}\land\vec{r}$, at most of the order
of $\Omega R_\perp$, is much smaller than the typical velocity
field in the lab frame
at a distance $\xi$ from the vortex core, $v\sim \hbar/m\xi$,
in the Thomas-Fermi limit, so that the structure of the vortex line
is not distorted by rotation. Another consequence is that 
the envelope function is also not destabilized by the rotation
\cite{personal_estimate} and is close to the usual Thomas-Fermi expression.

Expressing the phase of the condensate as function of the vortex line
is made difficult by the spatial inhomogeneity of the density profile
of the condensate \cite{pas_uni}. In principle this phase has to be determined 
everywhere in the condensate if one wants to calculate the kinetic energy
term of (\ref{eq:ener}). Fortunately, using the fact that the
condensate is in a steady state, one can replace the volume integral 
giving the kinetic energy stored in the condensate phase by a
line integral along the vortex line, using the same type
of techniques as in \cite{AftaRiv}. It is then possible
to rely on approximations for the condensate phase valid close to the 
vortex line. We use in the appendix \ref{appendix} the simplifying hypothesis
that the vortex line is weakly curved, with a radius of curvature
of the order of $R_\perp$ or larger, which allows to approximate
the condensate phase close to the vortex line by the one of a 
straight vortex.

We finally obtain the following energy functional of the vortex
line, taking the vortex free configuration as the zero of energy:
\begin{eqnarray}
E_{\rm v} \simeq 
\int ds &&\frac{g_{\rm 2D}(z_0(s))}{g}
\left[E_{\rm 2D}^{\Omega=0}(r_{0\perp}(s);z_0(s)) \right.\nonumber \\
&&\left.+ \cos(\alpha(s)) E_{\rm 2D}^{\rm rot}(r_{0\perp}(s);z_0(s))\right].
\label{eq:final_main}
\end{eqnarray}
In this expression the vortex line is parametrized by the curvilinear abscissa 
$s$. At the point of abscissa $s$ the vortex line is at the elevation $z_0(s)$
and at a distance $r_{0\perp}(s)$ from the rotation axis, and makes an
angle $\alpha(s)$ with respect to $z$.
A remarkable feature of (\ref{eq:final_main})
is that it is expressed in terms of the energy functionals
of a vortex core in a 2D condensate, $E_{\rm 2D}^{\Omega=0}$ for the energy
in the absence of rotation and $E_{\rm 2D}^{\rm rot}$ for the energy due
to the $-\Omega L_z$ term.
This is physically plausible considering the cigar shaped nature of the condensate, 
and this allows to view the 3D condensate as a collection of 2D horizontal
slices.  The slice of elevation $z$ constitutes a 2D Bose condensate with
a chemical potential $\mu-m\omega_z^2z^2/2$, where $\mu$ is the chemical potential
of the 3D condensate, and has a Thomas-Fermi radius
coinciding with the local 3D one. 
The coupling constant $g_{\rm 2D}(z)$ of the 2D gas 
can be expressed in terms of the 3D coupling constant, see (\ref{eq:g_2D}). We
arrive at the simple formula
\begin{equation}
\frac{g_{\rm 2D}(z)}{g} = \frac{15}{16R_z}\left(1-\frac{z^2}{R_z^2}\right)^2
\end{equation}
where $R_z$ is the Thomas-Fermi radius of the condensate along $z$.
An interesting remark is that the rotational energy term in (\ref{eq:final_main}) 
is multiplied by $\cos\alpha(s)$. As $E_{\rm 2D}^{\rm rot}$ is proportional to the
rotation frequency $\Omega$, this means that $\cos\alpha(s)E_{\rm 2D}^{\rm rot}$
is the rotational energy of a vortex core in a 2D condensate rotating at the
effective frequency
\begin{equation}
\Omega_{\rm 2D}(s) = \Omega \cos\alpha(s).
\end{equation}

\subsection{Minimizing numerically the simple energy functional} 
In a first stage we have to check that the energy functional derived
in the appendix \ref{appendix} correctly reproduces the results of the
minimization of the full Gross-Pitaevskii energy functional.
We perform this check numerically:
we discretize the vortex line in
little segments having all the same length $dl$ much smaller than
the transverse Thomas-Fermi radius $R_\perp$ of the condensate. 
As the vortex line is symmetric with respect to $z$ reflexion and lies
in the $xz$ plane,  the left extremity of the
first segment in the calculation moves along $x$ axis only, with
an abscissa $x_0$. The $z>0$ part of the vortex line
is discretized in $k$ segments and its shape is parametrized by the
angles $\alpha_i$, $i=1,\ldots,k$ at which the $k$ segments 
are with respect to 
the axis $z$. The energy functional (\ref{eq:final_main}) in its discretized version
is then a function of $k+1$ coordinates, 
that is of $x_0$ and of the $k$ angles $\alpha_i$. 
Starting with
a straight vortex line
at some small angle with respect to $z$ axis we move $x_0$
and the $\alpha_i$'s according to the simple gradient method or imaginary
time evolution method, that is we move the parameters by a small step
in the direction opposite to the local gradient of the energy
functional. 

It is known that this simple gradient method is not efficient when
the desired minima are at the bottom of a very elongated valley in the coordinate
space \cite{numrec}. This potential problem is minimized by a rescaling
of the coordinates by their
natural units $x_0^{\rm typ}=R_\perp$ and $\alpha^{\rm typ}=2\pi$,
so that our specific algorithm is to iterate the following small coordinate changes 
\begin{eqnarray}
dx_0 &=& -d\tau \left(x_0^{\rm typ}\right)^2
\partial_{x_0} E_{\rm v}/E_v^{\rm typ} \\
d\alpha_i &=& -d\tau \left(\alpha^{\rm typ}\right)^2
\partial_{\alpha_i} E_{\rm v} /E_v^{\rm typ}
\end{eqnarray}
where the typical energy scale is $E_v^{\rm typ}=\hbar^2\omega_\perp^2/\mu$ 
and the dimensionless \lq time' step $d\tau$ 
is small as compared to unity (typically $0.1$).
The iteration stops when $|dx_0|/(d\tau x_0^{\rm typ})$
and $|d\alpha_i|/(d\tau \alpha^{\rm typ})$ are below some threshold, 
taken here to be $10^{-6}$.

\begin{figure}
\centerline{\includegraphics[width=8cm,clip=]{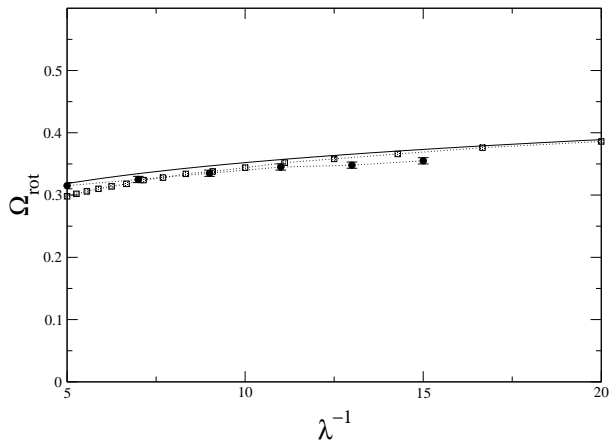}}
\caption{Minimal rotation frequency required to stabilize a vortex in a cigar-shaped
condensate, as function of $\lambda^{-1}=\omega_\perp/\omega_z$. 
The parameters are given in the last paragraph of 
section \ref{sec.model}. Disks with error bars: minimization of the full
Gross-Pitaevskii energy functional. Squares: minimization of the approximate 
vortex line energy functional, based on the 2D $\eta$-modified energy functional
(\ref{eq:eta_in}) and a discretization of the $z>0$ part
of the vortex line in $k=256$ segments. 
Solid line: analytical estimate of subsection \ref{subsec:enfin}.
\label{fig:line_mini}}
\end{figure}

The resulting prediction for the minimum rotation frequency required to
stabilize the vortex line is shown in Figure \ref{fig:line_mini},
deliberately restricted to the domain of cigar shaped condensates.
The agreement with the minimization of the full Gross-Pitaevskii energy functional
is remarkable, considering the fact that the points of Figure
\ref{fig:line_mini} are moderately in the Thomas-Fermi regime
($\mu\sim 7.6\hbar\omega_\perp$ for $\omega_\perp/\omega_z=15$).

\subsection{Why the vortex line bends in a cigar shaped condensate}
\label{subsec:enfin}
\begin{figure}
\centerline{\includegraphics[width=8cm,clip=]{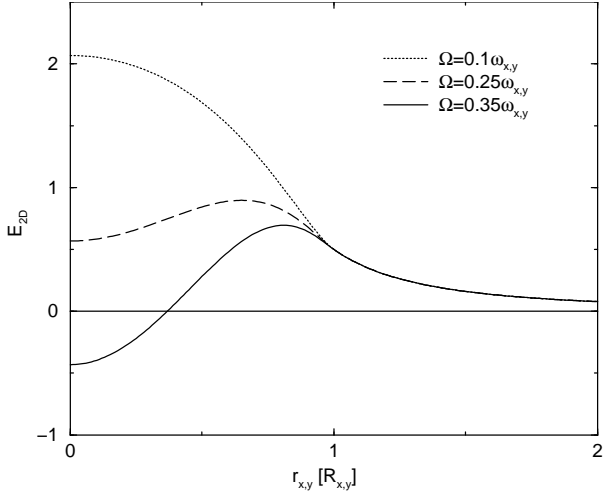}}
\caption{Energy of a single vortex in a 2D Thomas-Fermi condensate in a harmonic trap
$m\omega_\perp^2 (x^2+y^2)/2$
rotating at frequency $\Omega$, as function 
of the distance of the vortex core to the trap center, for a chemical
potential $\mu_{\rm 2D}=10\hbar\omega_\perp$. Dotted line: $\Omega=0.1\omega_\perp$.
Dashed line: $\Omega=0.25\omega_\perp$. Solid line: $\Omega=0.35\omega_\perp$. 
The values of the stabilization and critical rotation frequencies defined in the text,
see (\ref{eq:omega_sta_eta}) and (\ref{eq:omega_c_eta}),
are $\Omega_{\rm stab}^{\rm 2D}\simeq 0.170\omega_\perp$ and
$\Omega_{\rm c}^{\rm 2D}\simeq 0.306\omega_\perp$.
The energy is in units of $\hbar^2\omega_\perp^2/\mu_{\rm 2D}$
where $\mu_{\rm 2D}$ is the chemical potential of the 2D gas,
and is calculated from the approximate formulas (\ref{eq:ener_rot},
\ref{eq:eta_out},\ref{eq:eta_in}).
The distance is in units of the Thomas-Fermi radius of the 2D condensate.
\label{fig:case}}
\end{figure}

The actual goal of this section is to understand physically why the vortex
line bends in a cigar shaped condensate.
This can be achieved intuitively thanks to the very suggestive
form of the energy functional (\ref{eq:final_main}).
One just needs to have in mind the following characteristics of the 2D vortex problem:
\begin{itemize}
\item if the effective rotation frequency $\Omega_{\rm 2D}$ is too small
the 2D energy functional has a maximum for the vortex core at the center
of the trap and is a purely repulsive potential, see the dotted line in
figure \ref{fig:case}: the vortex core cannot be
stabilized inside the condensate and its equilibrium position is at infinity.
\item if $\Omega_{\rm 2D}$ is above the stabilization frequency $\Omega_{\rm stab}^{\rm 2D}$
and below the critical rotation frequency $\Omega_{\rm c}^{\rm 2D}$, the 2D energy functional
has a {\it local} but not global 
minimum for the vortex core at the center of the trap (dashed line
in Figure \ref{fig:case}). In this situation,
the vortex core
is stabilized at the trap center with an energy larger than the vortex
free condensate.
\item for $\Omega_{\rm 2D}>\Omega_{\rm c}^{\rm 2D}$, 
the energy minimum at the trap center
is now below the energy of the vortex free configuration
(solid line in Figure \ref{fig:case}).
\end{itemize}
The important feature of the 2D case is that the equilibrium positions of the
vortex core are either the trap center or the infinity. Another point,
crucial for the 3D case, is that both $\Omega_{\rm stab}^{\rm 2D}$ and 
$\Omega_{\rm c}^{\rm 2D}$ are decreasing functions of the chemical
potential $\mu_{\rm 2D}$.

\begin{figure}
\centerline{\includegraphics[width=4cm,clip=]{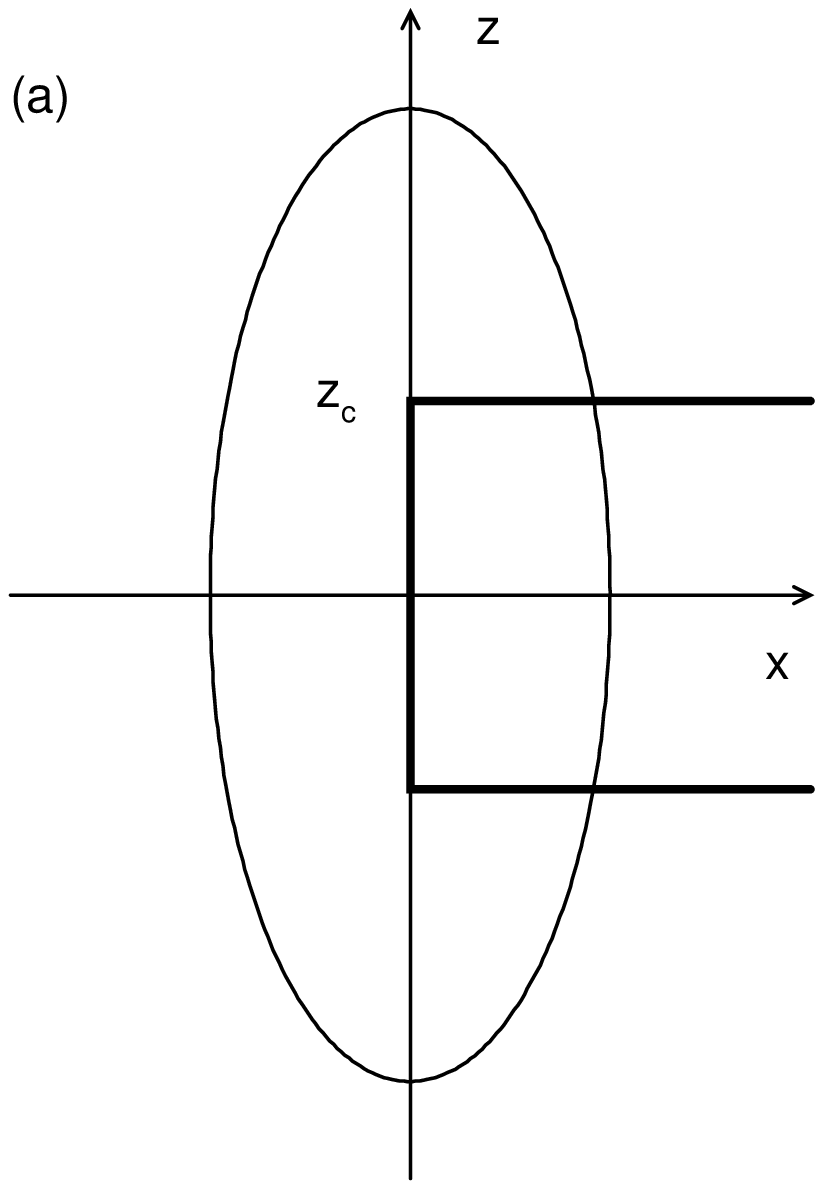}
\hspace{2mm}
\includegraphics[width=4cm,clip=]{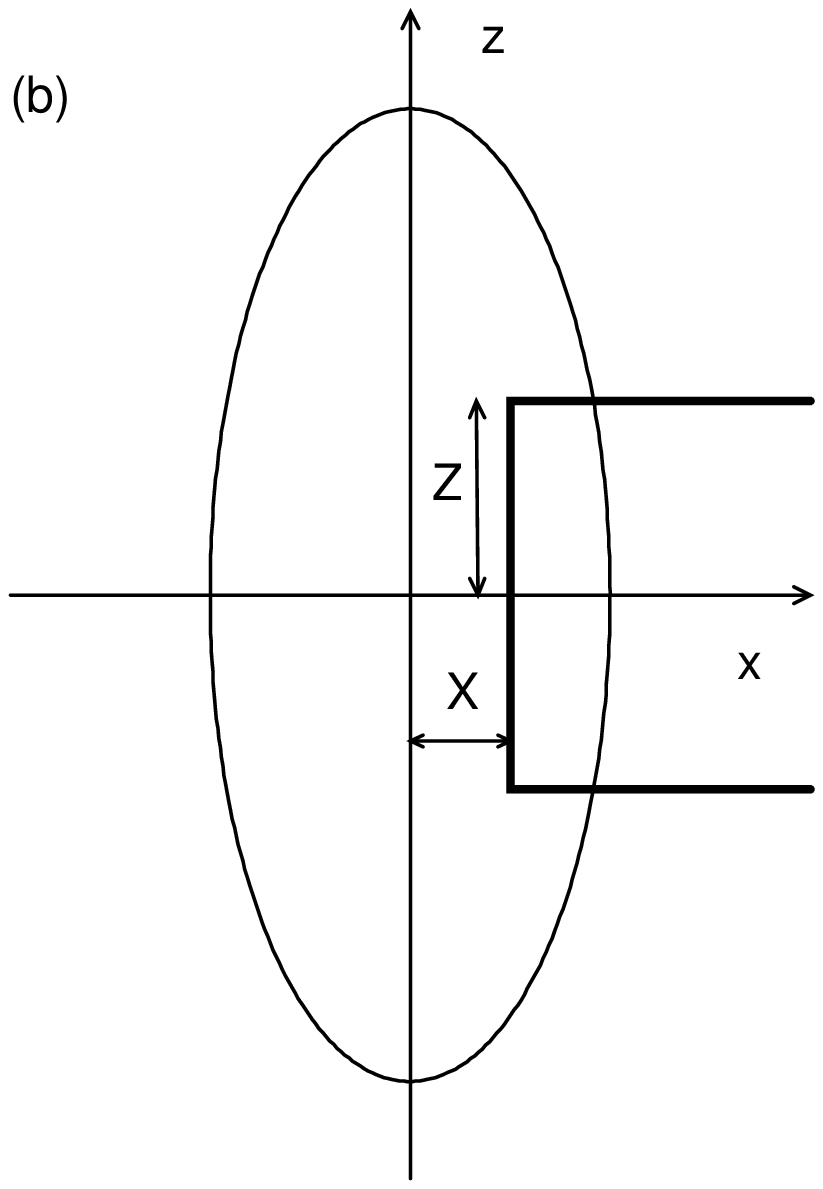}}
\caption{The two C-shaped ansatz used in this paper 
for the vortex line in a cigar-shaped condensate
in a trap rotating at frequency $\Omega$.
(a) the vortex line (thick solid line)
is on the rotation axis for an elevation in between
$-z_c$ and $z_c$, otherwise goes to infinity horizontally;
$z_c$ is the positive elevation of the 2D slice having a critical rotation frequency
$\Omega_{\rm c}^{\rm 2D}$ equal to $\Omega$.
(b) the C shape is at a distance $X$ to the rotation axis
and has a total height 2$Z$; $X$ and $Z$ are determined variationally.
\label{fig:petit_schema}}
\end{figure}

Let us now follow the vortex line travelling through the 3D cigar shaped condensate,
in the case where the 2D critical rotation frequency in the central slice $z=0$
is smaller than the actual rotation frequency $\Omega$. 
Let us call $z_{\rm c}$ the elevation of the 2D slice with a
local critical frequency $\Omega_{\rm c}^{\rm 2D}$ equal to $\Omega$, 
see Figure~\ref{fig:petit_schema}a. 

It is clear that the vortex line will be straight along the rotation axis
for $|z|<z_{\rm c}$: the vortex line is there in a valley corresponding
to the global minimum of energy of each local 2D slice.
When the vortex line reaches the domain of elevation $z>z_{\rm c}$, having a vortex core 
on the rotation axis costs more energy than having the vortex core at infinity in
each local 2D slice. The tempting strategy then offered to the vortex line
is to bend and leave the condensate. 
Assume that the vortex line leaves the condensate
radially, as shown in Figure~\ref{fig:petit_schema}a. The corresponding horizontal
vortex line has an energy
\begin{equation}
E_{\rm v}^{\rm horiz} = \frac{g_{\rm 2D}(z_c)}{g}
\int_0^{+\infty} dx\, E_{\rm 2D}^{\Omega=0}(x;z_c).
\end{equation}
The corresponding integral can be calculated exactly, but it is sufficient
to give here an order of magnitude: $g_{\rm 2D}/g$ scales as $1/R_z$,
$E_{\rm 2D}^{\Omega=0}$ is of the order of $\hbar^2\omega_\perp^2/\mu_{\rm 2D}$
and the integral over $x$ converges over a distance given by the Thomas-Fermi radius $R_\perp$,
so that
\begin{equation}
E_{\rm v}^{\rm horiz} \sim \frac{R_\perp}{R_z} \frac{(\hbar\omega_\perp)^2}{\mu}
\left(1-\frac{z_c}{R_z}\right)^{3/2}.
\end{equation}
We have included some approximate $z_c$ dependence relevant for the extreme
case of $z_c$ close to $R_z$.
What would be the energy cost for the vortex line to remain on the rotation
axis from $z_c$ to $R_z$? The energy of the corresponding vertical segment
is
\begin{eqnarray}
\label{eq:e_vert}
E_{\rm v}^{\rm vert} &=& \int_{z_c}^{R_z} dz\, \frac{g_{\rm 2D}(z)}{g}
\left[E_{\rm 2D}^{\Omega=0}(0;z) + E_{\rm 2D}^{\rm rot}(0;z)\right] \\
&=& \frac{15}{16 R_z} \int_{z_c}^{R_z} dz\, \left(1-\frac{z^2}{R_z^2}\right)^2
\left[\hbar\Omega_{\rm c}^{2D}(z)-\hbar\Omega_{\rm c}^{2D}(z_c)\right]\nonumber
\end{eqnarray}
where we have used the fact that the 2D rotational energy of a vortex core
in the center of a trap rotating at frequency $\Omega$ is $-\hbar\Omega$,
also equal to $-\hbar\Omega_{\rm c}^{2D}(z_c)$ by definition of $z_c$.
The integrand has the
same order of magnitude as in $E_{\rm v}^{\rm horiz}$ but the integration
length is now of the order of $R_z$ so that
\begin{equation}
E_{\rm v}^{\rm vert} \sim \frac{\hbar^2\omega_\perp^2}{\mu} 
\left(1-\frac{z_c}{R_z}\right)^2
\end{equation}
is typically $R_z/R_\perp$ times larger than $E_{\rm v}^{\rm horiz}$.
This proves that in the limit of a cigar-shaped
condensate $R_z\gg R_\perp$, the strategy of bending is more favorable
than the strategy of following the rotation axis, except
for a $z_c$ very close to the end point of the condensate,
$1-z_c/R_z \sim R_\perp^2/R_z$
that is $\Omega\sim (\hbar^2\omega_\perp^2/\mu)(R_z/R_\perp)^2$.

The bending at the value $z_c$ in the above reasoning can be
justified variationally as follows.
We perform a simple
minimization of the energy functional (\ref{eq:final_main}) with the following
linear piecewise variational ansatz for the vortex line in the plane
$y=0$:
an horizontal line linking $(x=+\infty,z=-Z)$ to $(x=0,z=-Z)$,
then a vertical segment of length $2Z$ along the rotation axis,
then an horizontal line from $(x=0,z=Z)$ to $(x=+\infty,z=Z)$. The energy
of the ansatz depends on the single variational parameter $Z$:
\begin{equation}
E_{\rm v}^{\rm ans}(Z) = E_{\rm v}^{\rm segm}(Z)
+E_{\rm v}^{\rm line}(Z)
\end{equation}
that is the sum of the energies of the vertical segment and of the
horizontal lines.
One has to extremize this function over $Z$. In the case where the extremum
is in the interior of the interval $(0,R_z)$, one has to solve
\begin{equation}
\frac{d}{dZ} E_{\rm v}^{\rm ans} = 0.
\label{eq:extreme}
\end{equation}
This non-trivial task becomes simple in the limit of a very
elongated condensate along $z$ \cite{proviso}.
As shown in appendix \ref{appendix}, the vortex 2D energy function 
$E_{\rm 2D}^{\Omega=0}(r_\perp;z)$ depends on $r_\perp$ only {\it via}
the ratio $r_\perp/R_\perp(z)$ where $R_\perp(z)$ is the local Thomas-Fermi
radius. As a consequence $E_{\rm v}^{\rm line}(Z)$ depends on $Z$
through $Z/R_z$ only, and its derivative is 
$R_\perp/R_z$ times smaller
than the derivative of $E_{\rm v}^{\rm segm}(Z)$. In the limit
$R_z/R_\perp$ tending to infinity, equation (\ref{eq:extreme}) reduces to
\begin{equation}
\frac{d}{dZ} E_{\rm v}^{\rm segm}(Z)  = 0 .
\end{equation}
Using the explicit expression of $E_{\rm v}^{\rm segm}(Z)$
similar to (\ref{eq:e_vert}) we obtain the condition
\begin{equation}
\Omega_{\rm c}^{\rm 2D}(Z) = \Omega
\end{equation}
that is $Z=z_{\rm c}$,
the vortex starts bending at the elevation where 
the 2D critical rotation frequency equals the trap rotation frequency $\Omega$.
As a consequence the minimal rotation frequency to stabilize the bent vortex line
is given in the $\omega_z/\omega_\perp\to 0$ limit by the 2D critical
frequency in the central slice $z=0$, see (\ref{eq:omega_c_eta}):
\begin{equation}
\Omega_1 =
\frac{\hbar\omega_\perp^2}{\mu}
\log\left[e^{C+1/2}\left(\frac{\mu}{\hbar\omega_\perp}
+\eta\right)
\right]
\label{eq:fini}
\end{equation}
with $\eta\simeq 1.938$ and $C\simeq 0.0884$. This asymptotic prediction 
is plotted as a solid line in Figure \ref{fig:line_mini} and is in good agreement
both with the numerical minimization of the vortex energy functional and
with the numerical minimization of the full Gross-Pitaevskii energy functional
\cite{general}.

To conclude this subsection
we point out two striking properties of the result (\ref{eq:fini}).
Firstly, it explains why in a cigar-shaped condensate,
the bent vortex is first stabilized with an almost vanishing energy gap
with respect to the 0-vortex configuration, at least  much smaller than
$\hbar^2\omega_\perp^2/\mu$, see Figure \ref{fig:lz}:
at $\Omega=\Omega_1$, the variational ansatz predicts
an energy scaling as $(\hbar^2\omega_\perp^2/\mu)R_\perp/R_z$,
see the expression for $E_v^{\rm horiz}$ with $z_c/R_z \simeq 0$. This is
very different from the 2D case, where the vortex, when first stabilized,
has a large and positive energy $\sim \hbar^2\omega_\perp^2/\mu$
with respect to the vortex free configuration.
Secondly, it is remarkable that the 2D stabilization frequency plays no role
in the 3D case for the cigar-shaped condensate. 
This means that being a local minimum
of energy in 2D slices does not imply that the vortex line is a local minimum of
energy in 3D. As an example, it is possible in 3D to shorten the
vertical segment of the vortex line, that is to introduce the bending at a
slightly lower elevation, whereas this infinitesimal
transformation has no equivalent in a purely 2D case.

\subsection{Saddle points of the simple energy functional}

Using a simple energy functional rather than the full Gross-Pitaevskii
energy functional has allowed to understand the bending of the vortex line
physically with the help of a simple piecewise variational ansatz.
But much more can be done: as we now show, one can investigate 
not only the energy minimum but also possible saddle points of the
energy functional. This will explain an intriguing feature of
the numerical results of Figure \ref{fig:lz}:
the angular momentum of the condensate has a discontinuous jump
from the one-vortex branch to the no-vortex branch, which opens
a gap in the allowed values of $\langle L_z\rangle$.

The idea is to consider now a more general ansatz for the vortex
line than the one of the previous subsection, in order to allow
the vortex line to move towards the border of the condensate. 
We introduce the two parameter linear piecewise ansatz in the $y=0$ plane
shown in Figure~\ref{fig:petit_schema}b:
a horizontal line linking $(x=+\infty,z=-Z)$ to $(x=X,z=-Z)$,
then a vertical segment of length $2Z$ parallel to $z$, and
finally a horizontal line from $(x=X,z=Z)$ to $(x=+\infty,z=Z)$.
In other words, the vortex line has a C shape of adjustable distance $X$
to the rotation axis and adjustable height $2Z$.

\begin{figure}
\centerline{\includegraphics[width=8cm,clip=]{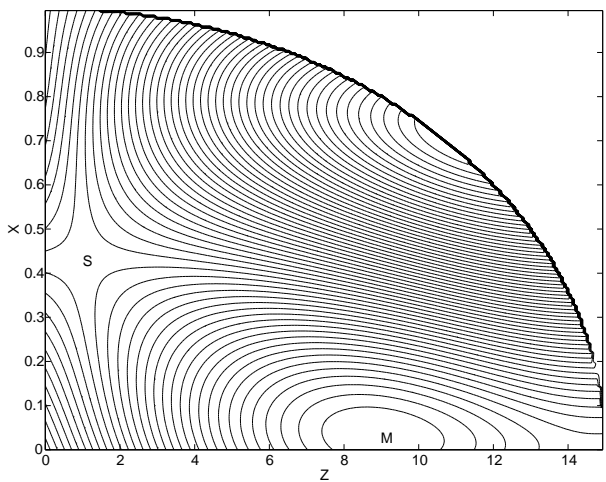}}
\caption{For the two parameter linear piecewise ansatz for the vortex line
shape, contour plot of the simple energy functional as function
of the coordinates  $(X,Z)$ of the bending point of the line.
The trap aspect ratio is 
$\lambda=\omega_z/\omega_\perp=1/15$, leading to a Thomas-Fermi
chemical potential $\mu=7.62\hbar\omega_\perp$. The rotation frequency
is $\Omega=0.448\omega_\perp\simeq 1.2\Omega_1$.
$X$ and $Z$ are expressed in units of the transverse Thomas-Fermi
radius $R_\perp$ of the condensate.
The contour plot is restricted to a bending point inside the Thomas-Fermi
profile of the condensate, $X^2+\lambda^2 Z^2 < R_\perp^2$.
The global minimum is marked with the letter M, and the saddle
point is marked with the letter S. The elliptic isocontours
close to the Thomas-Fermi border correspond to a maximum of energy.
\label{fig:contour}}
\end{figure}

We consider the parameters of the ENS experiment, 
with $\lambda=\omega_z/\omega_\perp =1/15$. Eq.(\ref{eq:mu_explicit})
leads to a Thomas-Fermi chemical potential of $\mu=7.62\hbar\omega_\perp$.
We choose a rotation frequency $\Omega=0.448\omega_\perp$,
which is a factor 1.2 above the threshold value $\Omega_1$
of (\ref{eq:fini}). Figure \ref{fig:contour} shows
a contour plot of the simple energy functional as function
of the coordinates $(X,Z)$ of the bending point of the vortex
line. One recognizes first the global minimum of energy, marked with
a M, with a position of the C shaped vortex line similar to the one of
Figure~\ref{fig:petit_schema}a: the vertical part of the
vortex line is very close to the rotation axis, with $X$ less than
0.05 times the transverse Thomas-Fermi radius $R_\perp$.
Second, one finds a saddle point in the energy, marked with a S, corresponding
to a quite different position of the C shape: 
the vortex line is now far from the rotation
axis, with $X\sim 0.45 R_\perp$, and the half elevation of the
$C$ shape is of the order of $R_\perp$.

What is the physical meaning of this saddle point~?
As it corresponds to an extremum of the energy functional (vanishing
first order derivatives of the functional), it represents
a stationary shape for the vortex line in the rotating frame.
In addition to this dynamic property, it has the following
interesting energetic aspect: it corresponds to an energy minimum
for a fixed value of the angular momentum per particle
$\langle L_z\rangle$, as one can check explicitly for the
two-parameter ansatz.

These properties are not specific
to the two-parameter ansatz. They apply for arbitrary shapes
of the vortex line, but also for the exact Gross-Pitaevskii energy
functional. Consider indeed the subspace of condensate
wavefunctions with a fixed angular momentum per particle
$l_z$. The Gross-Pitaevskii
energy functional in the absence of rotational term
$-\Omega L_z$ is bounded from below (for $g>0$) in this subspace
so it has a minimum for some wavefunction $\psi_{l_z}$, 
with an energy $E^{\Omega=0}(l_z)$. One can then show that the wavefunction
$\psi_{l_z}$ is a stationary point for the full Gross-Pitaevskii energy
functional, that is including the rotational energy term $-\Omega L_z$,
without imposing a fixed the angular momentum, provided
that one takes the rotation frequency equal to $\Omega=dE^{\Omega=0}/dl_z$. 
One can show that
it is however not necessarily a minimum of the full energy functional: 
it is a minimum if 
\begin{equation}
\label{eq:thermo}
\frac{d\Omega}{dl_z}\equiv \frac{d^2E^{\Omega=0}}{dl_z^2}>0
\end{equation}
otherwise it is a saddle point.

\begin{figure}
\centerline{\includegraphics[width=8cm,clip=]{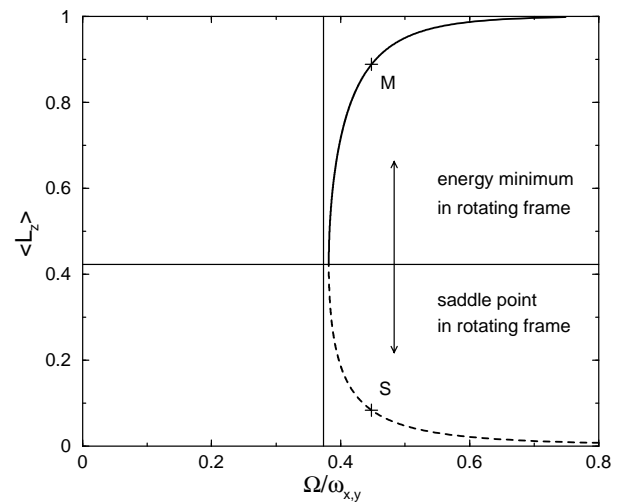}}
\caption{For the two-parameter ansatz and the simple energy functional
for the bent vortex line, mean angular momentum per particle 
for the energy minimum (solid line)
and the energy saddle point (dashed line)
as function of the rotation
frequency $\Omega$. The trap aspect ratio is $\lambda=1/15$ and the chemical
potential is $\mu=7.62\hbar \omega_\perp$. The points marked with M and
S correspond to the rotation frequency of Figure
\ref{fig:contour}. The vertical line corresponds to the analytical
prediction for $\Omega_1$. The rotation frequencies are in 
units of $\omega_\perp$ and the angular momentum is in units
of $\hbar$.
\label{fig:lz_vs_rot}}
\end{figure}

We can exemplify these properties with our simple two-parameter ansatz.
We have plotted in Figure~\ref{fig:lz_vs_rot} the mean angular momentum
per particle of the energy minimum and of the saddle point as function 
of the rotation frequency $\Omega$. One finds that the branch of minimum
has an angular momentum ranging from 0.42$\hbar$ to $\hbar$, 
whereas the branch of saddle 
point fills the gap of angular momentum from 0 to 0.42$\hbar$.
One also finds that the angular momentum is an increasing function
of the rotation frequency on the branch of minimum, whereas it is
a decreasing function of $\Omega$ on the saddle point branch, in accordance
with the general condition (\ref{eq:thermo}).

In an experiment where the angular momentum of the condensate
is gradually decreased, starting from a value close to $\hbar$, the vortex
first follows the minimum energy branch: the distance of the vortex
line to the rotation axis $X$ remains much smaller than $R_\perp$, whereas
the vortex half-height $Z$ and the rotation frequency $\Omega$ 
of the line in the lab frame decrease with time. Then the angular
momentum reaches a critical value: the minimum in the $(X,Z)$
parameter space merges with the saddle point. When
the angular momentum is further decreased, the vortex line
follows the saddle point
branch: the distance $X$ increases and becomes of the order
of $R_\perp$, as the vortex moves towards the boundary
of the condensate; $Z$ takes values as small as $R_\perp$, in
which case the validity of our simple analytical approach
becomes marginal, and the rotation
frequency of the vortex line increases. This scenario
was very recently observed at ENS \cite{vu_a_ENS}.

To conclude this subsection, we can emphasize again the analogy
of the 3D bent vortex problem and the 2D single vortex problem. 
In 2D, the minimum
of energy in the rotating frame with no constrainst
on the angular momentum corresponds to the vortex at the trap
center, with an angular momentum $\hbar$, or to the vortex at infinity
with a vanishing angular momentum. If one minimizes the energy for
a fixed angular momentum $l_z$, one finds an off-center vortex core,
shifting from the trap center to infinity as $l_z$ is varied from $\hbar$
downto zero; these off-center solutions do not
satisfy the criterion (\ref{eq:thermo}) \cite{Rokhsar} and are
very analogous to our saddle point solutions.

\section{Finite temperature fluctuations of an otherwise straight vortex line}
\label{sec:finite_temp}

\subsection{Method: Bogoliubov approach around the straight vortex
steady state}

In order to evaluate the contribution of thermal fluctuations
on the contrast, we study the limiting case of a straight
vortex. For the ENS trap parameters, this means that $1/\lambda < 7$ and in
all the following, we illustrate our discussion using the case $(\lambda
= 1/5, \Omega=0.5\omega_\perp)$. In this regime, the system has a rotational 
symmetry around
the $z$ axis so that the numerical problem is effectively 2D in cylindrical
coordinates. We use the $U(1)$-symmetry preserving
Bogoliubov approach described in \cite{U1}. In this way, the
problem of the spurious mode of the condensate is avoided. The Bose
field is expanded as 
\begin{equation}
\hat{\psi}(\vec{r}\,) = {\phi}(\vec{r}\,) \hat{a}_{\phi}+
\hat{A}_\phi
\sum_k  \hat{b}_k u_k(\vec{r}\,) + \hat{b}_k^\dagger v_k^*(\vec{r}\,).
\label{eq:expan}
\end{equation}
$\phi$ is the condensate wave function normalized to unity. 
The operator $\hat{a}_\phi$ is the annihilation operator
in the condensate mode and the almost unitary operator $\hat{A}_\phi=
(\hat{a}_\phi^\dagger \hat{a}_\phi+1)^{-1/2} \hat{a}_\phi$ 
gives the phase factor
of the field in the condensate mode \cite{Girardeau,Wigner}. The modal
functions $u_k,v_k$ are normalized like 
$\int d^3{\vec{r}} \ |u_k|^2-|v_k|^2 = 1$. 
They are obtained from the usual modal functions of the Bogoliubov-de Gennes 
equations after orthogonalization of $u_k$ and of $v_k$
with respect to $\phi$. The
Bose operators $\hat{b}_k$ annihilate
one quasi-particle in the mode $k$ but conserve the total number of particles.
The index $k=\{n,l,s\}$ denotes the quantum numbers of the mode linked
to the symmetry of the system: $s=1$ (or $-1$) for symmetric (or antisymmetric)
modes with respect to the plane $z=0$, $l\hbar$ is the angular momentum 
with respect to the condensate and the integer $n$ is the radial quantum
number. 

Concerning the transverse direction $x-y$, we expand $\phi,u_k,v_k$ on the
harmonic oscillator basis $\left\{\Phi_{\rm ho}^{m',n'}\right\}$ of pulsation
$\omega_\perp$ ($m'\hbar$ is the angular momentum and $n'$, the radial quantum
 number). For example, in this basis we have 
\begin{equation}
\phi(\vec{r}\,) = \sum_{n'=0}^\infty c_{n'}(z) \Phi_{\rm ho}^{m=1,n'}(x,y)\,.
\end{equation}

Numerically, the basis is truncated: the spatial
grid  along $z$ is surrounded by infinite walls, and also the 
number of wave functions $\Phi_{\rm ho}^{m',n'}$ is limited  (for the value 
$\lambda=1/5$, the harmonic oscillator basis in our computation contains 
all the wave functions of energy less than $42$~$\hbar\omega_\perp$). The choice of the 
grid and the discretization of the Laplacian along $z$ has been made so 
that the first 160 energy levels $\epsilon_n$ of the pure 1D harmonic 
oscillator are recovered with an error $|\epsilon_n-\epsilon^{\rm exact}_n| 
<10^{-2}$. 
In this part we have computed the condensate wave function using an
imaginary time method and the convergence criterion (\ref{eq:criterion}) (typical
values of $\gamma$ are of the order of $10^{-13}$).

\subsection{Expectation values of some observables}

\begin{figure}
\centerline{\includegraphics[width=8cm,clip=]{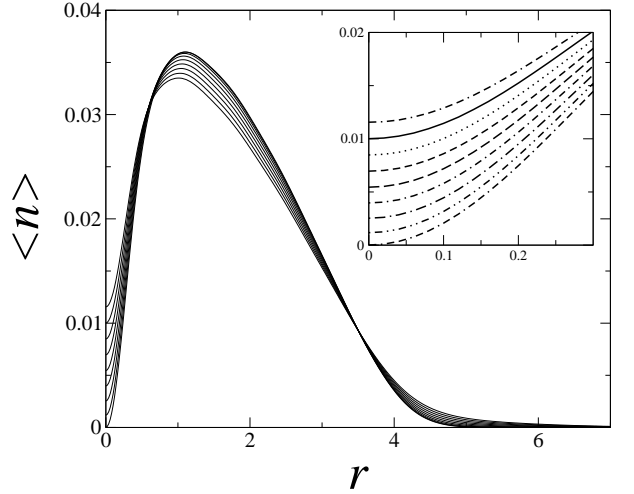}}
\caption{Mean column density $\langle n\rangle$  associated to 
the parameters $(\lambda=1/5,\Omega=0.5\omega_\perp)$, as function
of the distance $r$ to the rotation axis.
The temperature $T$ ranges from $0$ to $0.4T_c$
in steps of 0.05 $T_c$, from bottom to top. 
$r$ is in units of the harmonic oscillator 
length $a_\perp=(\hbar/m\omega_\perp)^{1/2}$ and $\langle n \rangle$ is in units of $Na_\perp^{-2}=Nm\omega_\perp/\hbar $. The number of particles is
$N=1.4\times 10^5$ and the chemical potential is $\mu\simeq 
9.87\hbar\omega_\perp$. The inset is a magnification.}
\label{fig:profil_T}
 \end{figure}

The mean density of atoms out of the condensate is obtained 
straightforwardly with the usual expression 
\begin{equation}
\rho_{\rm exc}(\vec{r}\,) =  \sum_{k} |v_{k}|^2(\vec{r}\,) + \sum_{k} n_{k}
\left( |u_{k}|^2(\vec{r}\,) + |v_{k}|^2(\vec{r}\,) \right) \, ,
\label{eq:dens_exc}
\end{equation}
where 
\begin{equation}
n_k=1/(\exp{(\epsilon_k/k_B T)}-1)
\label{eq:nk}
\end{equation}
is the Bose occupation factor and $\epsilon_k$ is the energy
of the Bogoliubov mode $k$ in the rotating frame. 
Because of numerical constraints we have limited the sum over
states of energy less than $20$~$\hbar\omega_\perp$ in the rotating frame, and we
have studied configurations with temperatures less than 
$11.4$~$\hbar\omega_\perp/k_B$ 
that is less than $0.4 T_c$, where $\displaystyle k_B T_c= \hbar \left(
{\omega_z\omega_\perp^2N}/{\zeta(3)}\right)^{1/3}$ 
is the ideal Bose gas critical temperature.
For comparison with experimental results, we extract the mean 
column density $\langle n\rangle$ in the $x-y$ plane:
\begin{equation}
\langle n\rangle(x,y) = \int dz \left[(N-\langle N_{\rm exc}\rangle)
|\phi|^2(\vec{r}\,)+\rho_{\rm exc}(\vec{r}\,) \right] \, ,
\end{equation}
where  $\langle N_{\rm exc}\rangle = \int d^3r \rho_{\rm
exc}(\vec{r}\,)$ 
is the mean number of atoms out of the condensate. We plot this
column density as function of the distance to the $z$ axis for
different values of the temperature in Figure~\ref{fig:profil_T}.
At zero temperature the effect of quantum depletion at the center of the vortex line
is not observable on the scale of the figure.
Hence quantum fluctuations of the vortex line are
clearly not responsible for the low contrast measured in ENS experiments.

\begin{figure}
\centerline{\includegraphics[width=8cm,clip=]{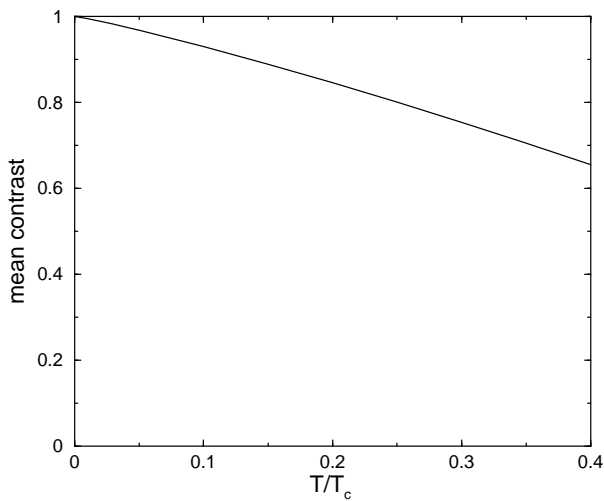}}
\caption{Dependence of the mean contrast $\overline{\cal C} = 1-\langle n\rangle^{\rm min} /
{\langle n\rangle^{\rm max}}$ with the
temperature for the same parameters as in Figure~\ref{fig:profil_T}. }
\label{fig:contraste}
\end{figure}

To be more quantitative, we denote $\langle n\rangle^{\rm min}$ the 
value of the mean column density at the center of the trap and 
$\langle n\rangle^{\rm max}$ the maximal value of the mean column density. 
Then, the mean contrast is defined as
\begin{equation}
\overline {\cal C} = 
1-\frac{\langle n\rangle^{\rm min}}{\langle n\rangle^{\rm max}} \,.
\end{equation}
The modal functions are of the form 
\begin{eqnarray}
&&u_k(\vec{r}\,) = U_k(r,z) \exp\left[i(l+1)\theta\right]\\
&&v_k(\vec{r}\,) = V_k(r,z) \exp\left[i(l-1)\theta\right] \quad ,
\end{eqnarray}
If $l+1 \neq 0$ the function $U_k$ vanishes in $r=0$ because of the centrifugal
barrier. For the same reason $V_k(r=0,z)$ vanishes for $l-1\neq 0$.
As a consequence only modes with $l=\pm 1$ 
contribute to $\langle n\rangle^{\rm min}$. Furthermore,
in the rotating frame the energy of the quasi-particles is 
given by Eq.(\ref{eq:E_omega}) and the most populated modes are the 
lowest energy kelvons, characterized by $l=-1$, which have a negative
energy in the absence of rotation \cite{Fetter_dynamics,feder}. 
In the example presented in this section, the lowest kelvon mode localized at
the extremities of the condensate has an energy in the rotating frame
given by
$E \simeq 0.018 \, \hbar \omega_\perp$. 
 Figure~\ref{fig:contraste} shows that  
the dependence of $\displaystyle \overline {\cal C}$ 
as a function of temperature is almost linear. 
This behavior can be understood from the fact that the lowest energy
modes have a semi-classical character ($\epsilon_k\ll k_B T$), 
with an occupation number depending linearly on
temperature $n_k \simeq k_B T/\epsilon_k$.

Figure~\ref{fig:contraste} shows that temperature contributes
significantly to the observed contrasts. In particular,
for a temperature of the order of the chemical potential,
which is a typical situation in experiments, we find
that the contrast is 65\%, not far from the contrast value
of $\sim$ 50\% observed at ENS.
Note that some zero temperature bending of the vortex line is expected 
to occur for the parameters of ENS, see Figure~\ref{fig:profil_michele},
which adds up to the thermal fluctuations and further
reduce the contrast in the experiment.

\subsection{Mimicking a single experimental run: a Glauber $P$ method}

Let us insist now on the fact that expectations values calculated
in the previous section are not sufficient for a quantitative
understanding of 
experiments. They just give an order of magnitude of what is
observed. Indeed, for example, for the particular value of $\lambda$ that we have
chosen, the straight vortex is at the edge of the existence domain and
one expects large fluctuations of the phase and of the density due to the soft
core mode and also to the emergence of low energy surface modes. As a
consequence, observables obtained from a single measurement of the
system may differ notably from their mean value. Hence, the aim of 
this section is to evaluate the fluctuations induced by the soft
modes. Calculations of the previous subsection
have shown that quantum fluctuations do
not contribute to the contrast at
finite temperature, so that semi-classical field approaches are good tools to
answer this problem. We use for that
purpose the Glauber $P$ method of quantum optics \cite{gardiner}.

This approach is made simple by the fact that
the $N$-body density operator $\hat{D}$ is
the exponential of $-H_{\rm Bog}/(k_B T)$, where the Bogoliubov Hamiltonian
is a sum of decoupled harmonic oscillator terms,
$H_{\rm Bog}=E_0 +\sum_k \epsilon_k \hat{b}_k^\dagger \hat{b}_k$.
We introduce the coherent state of quasi-particles $|\{\beta_k\}\rangle =
|\beta_{k_1},\beta_{k_2},\beta_{k_3}\ldots\rangle$, where $|\beta_k\rangle$ 
is the normalized eigenstate of the annihilation operators 
$\hat{b}_k$ with the eigenvalue $\beta_k$.
Then one defines the Glauber $P$ representation of  $\hat{D}$:
\begin{equation}
\hat{D} = \int \prod_k \frac{d \mbox{Re}\beta_k\,d\mbox{Im}\beta_k}{\pi}
P(\{\beta_{k'}\}) |\{\beta_{k'}\}\rangle \langle \{\beta_{k'}\} |\,.
\label{eq:density-op}
\end{equation}
The explicit calculation of $P$ is possible, as $D$ is Gaussian in
the $b_k$'s \cite{gardiner}, this gives the following 
Gaussian distribution:
\begin{equation}
P(\{\beta_{k'}\}) = \prod_k \frac{1}{n_k} \exp\left(-
\frac{\beta_k^*\beta_k}{n_k}\right) \,,
\end{equation}
where $n_k$ is the mean number of quasi-particles in the mode $k$
given in Eq.~(\ref{eq:nk}).
This Gaussian distribution is easily sampled. As $P$ is positive
one can imagine that a given experimental realization of the gas
is in the state $|\{\beta_k\}\rangle$ where the complex numbers
$\beta_k$ vary randomly from one realization to the other. In the
following, we want to determine
the density and the velocity field of the atomic gas
for a given Monte Carlo realization of the $\{\beta_k\}$. 

The $N$-body distribution function corresponding to
a single term of the statistical mixture (\ref{eq:density-op})
\begin{eqnarray}
\rho(\vec{r}_1, \ldots, \vec{r}_N) &= &
\langle \{\beta_{k'}\}|
\hat{\psi}^\dagger(\vec{r}_1\,)\ldots \hat{\psi}^\dagger(\vec{r}_N\,)
\nonumber \\
&&\times \hat{\psi}(\vec{r}_N\,)\ldots\hat{\psi}(\vec{r}_1\,) |\{\beta_{k'}\}\rangle
\label{eq:rhon}
\end{eqnarray}
is not easy to calculate for the interacting Bose gas as $\hat{\psi}$
is a superposition of $\hat{b}_k$ and $\hat{b}_k^\dagger$
so that the product of field operators in (\ref{eq:rhon})
is not normally ordered in terms of the $\hat{b}_k$.
We perform the following approximation
\begin{equation}
\hat{b}_k^\dagger |\{\beta_{k'}\}\rangle
\simeq \beta_k^* |\{\beta_{k'}\}\rangle.
\end{equation}
The error has a root mean square norm equal to unity, hence the
approximation is good for modes with a large occupation number, 
bad for empty
modes. In this way we describe correctly the fluctuations due to 
finite temperature,
but not the quantum fluctuations existing even at zero temperature.
As we have shown previously, this is fine in the dilute limit 
$(\rho a^3)^{1/2}\ll 1$, where
quantum depletion is small. The approximation amounts to taking
\begin{equation}
\hat{\psi}(\vec{r}\,) |\{\beta_{k'}\}\rangle \simeq \psi(\vec{r}\,) 
|\{\beta_{k'}\}\rangle
\end{equation}
with a classical field
\begin{equation}
\psi(\vec{r}\,) = \sqrt{N_0}\phi(\vec{r}\,) + \sum_k \beta_k 
u_k(\vec{r}\,) + \beta_k^*
v_k^*(\vec{r}\,) \, ,
\end{equation}
where $N_0$ is such that the norm squared of $\psi$ is equal
to the total number of particles $N$ \cite{phase_globale}.
In this approximation, a single realization is now in the coherent
state $|\{\psi\}\rangle$ for the field $\hat{\psi}$. In this case the $N$-body 
distribution function of the atoms for a single experimental 
realization is factorized:
\begin{equation}
\rho(\vec{r}_1,\ldots,\vec{r}_N\,)= \prod_k |\psi(\vec{r}_k\,)|^2 \, .
\end{equation}

\begin{figure}
\centerline{\includegraphics[width=8cm,clip=]{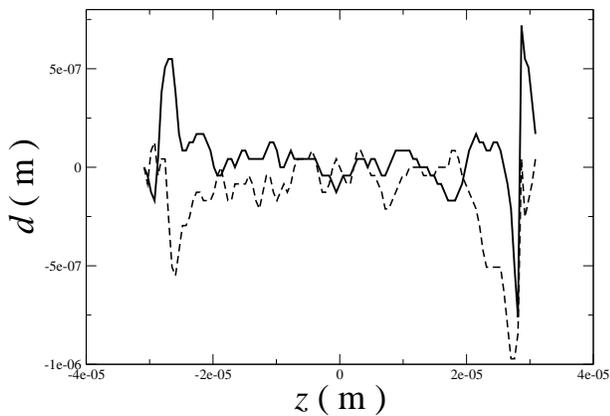}}
\caption{Representation of the $x$ coordinate (solid line)
and the $y$ coordinate (dashed line)
of the vortex line as function of the position $z$ on the rotation
axis, for a single stochastic realization.
The parameters are as in Figure~\ref{fig:profil_T}, with $T=0.2T_c$.
Note that the transverse Thomas-Fermi radius is of the order
of $6\mu$m.
}
\label{fig:line_d}
\end{figure}

\begin{figure}
\centerline{\includegraphics[width=9cm,height=9cm,clip=]{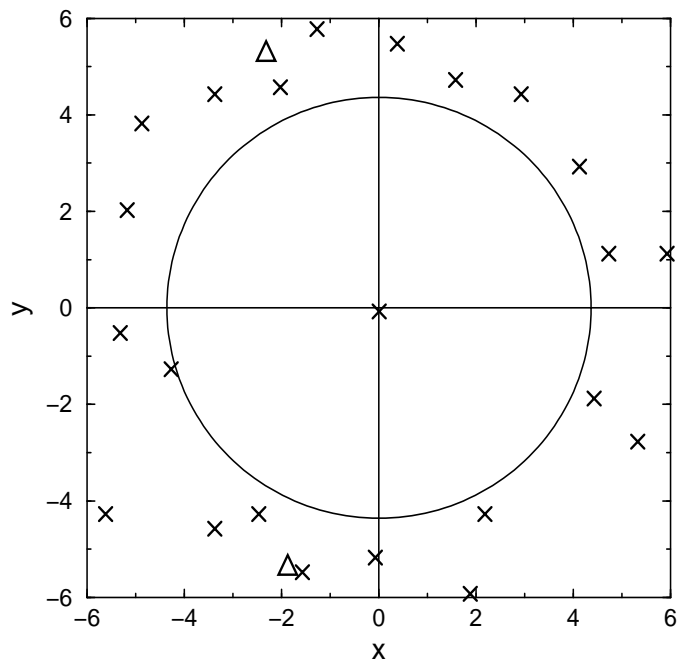}}
\caption{
For an individual Monte Carlo realization of the Bose
field $\psi$, position of the vortices in the velocity field
of the gas in the plane $z=0$, for
a temperature $T=0.2 T_c$. 
The vortices with a positive charge are indicated with 
a cross, the vortices with a negative charge are indicated
with a triangle. The circle indicates the Thomas-Fermi border
of the condensate. The $x$ and $y$ coordinates are in units
of $(\hbar/m\omega_\perp)^{1/2}$.
As a consequence of thermal excitation of the low
energy surface modes, satellite vortices appear at the border of the
condensate, around the central vortex. The physical parameters
are as in Figure~\ref{fig:profil_T}.}
\label{fig:satellite}
\end{figure}

In this paragraph, we describe temperature effects in a
single stochastic realization of $\psi$ at  $k_B T =0.2 k_B T_C \sim \frac{1}{2}\mu$.
First, we have extracted  the shape of the vortex line: 
Figure~\ref{fig:line_d} shows the $x$ and $y$ coordinates
of the vortex line as function of $z$.
The soft core modes localized at the two extremities
\cite{Fetter_dynamics,feder} of the condensate are 
thermally excited and as
a result push the core away from the rotation axis. This is a
temperature induced bending. 
Second, for the same
stochastic realization, Figure~\ref{fig:satellite}
represents the location of the vortices
in the velocity field of the gas in the plane $z=0$.
It shows that satellite vortices, which are not easily observable in
the density profile, appear at the border
of the condensate. The presence of these vortices is due to the
excitation of surface modes of low energy and relatively high angular
momentum (recall that for $\lambda=1/5,\Omega=0.5\omega_\perp$, 
the system is at
the border of the thermodynamic stability domain). 

\begin{figure}
\centerline{\includegraphics[width=8cm,clip=]{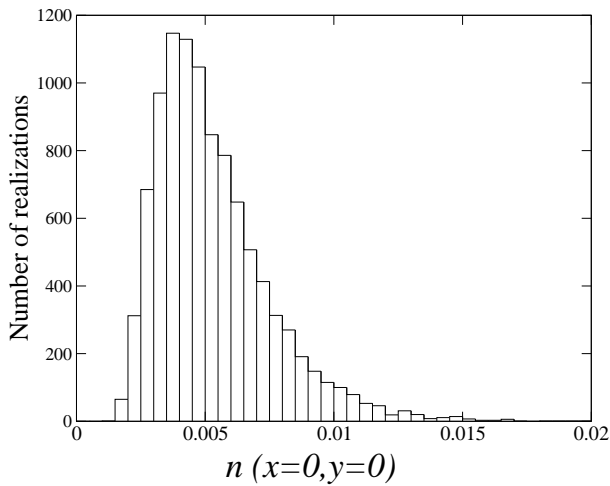}}
\caption{ Histogram associated to the single realization column density
$n$ at the center $x=0,y=0$, for $10^4$ realizations. 
This figure shows that the probability
law of this observable is strongly non Gaussian. $n$ is in units of 
$N/a_\perp^2= N m\omega_\perp/\hbar$.
The same parameters as in Figure~\ref{fig:profil_T} were used, with 
$T=0.2T_c$. We have observed that this non-Gaussian character
subsists even at a temperature as low as 0.05$T_c$.
}
\label{fig:coeurstat}
\end{figure}

We address now the problem of the integrated density
contrast for a single stochastic realization. In
the previous single stochastic realization at a temperature
$T=0.2 T_c$, the value of the contrast is around  $90\%$
which does not coincide with the mean contrast $\overline {\cal C}\simeq 85\%$
derived in the previous subsection.
 This disagreement
suggests strong fluctuations of density along the $z-$axis. To
confirm this idea we have computed 10000 stochastic realizations of the column
density at the center of the trap. We have reported in
Figure~\ref{fig:coeurstat} the corresponding  histogram.
 This figure shows effectively that the probability law is far from
being Gaussian: the long tail on the right side is due to 1D character of 
the excitation modes of the vortex line, the so-called kelvons,
similar indeed to the fluctuations of the number of condensate particles in
a 1D Bose gas \cite{cartago}. This is
another indication of the importance of fluctuations in the properties of this
system.

\begin{figure}
\centerline{\includegraphics[width=8cm,clip=]{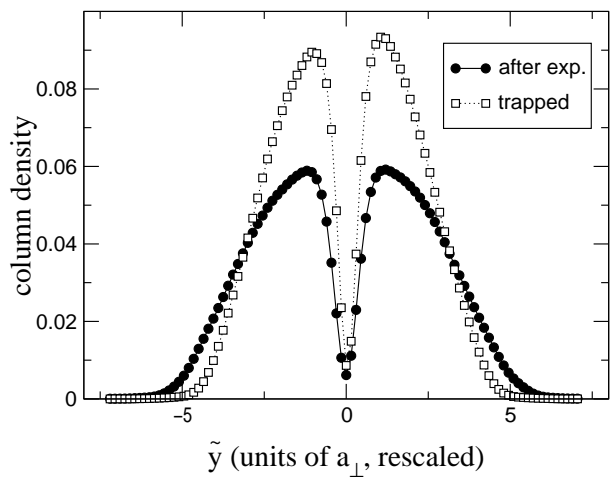}}
\caption{Comparison of the column density for a straight vortex 
($\lambda=1/5$) at finite temperature ($T=5.71~\hbar\omega_\perp/k_B$)
before and after the release from the trap
(time of flight $\sim30$ms), as a function of
the rescaled $\tilde{y}$ coordinate.
The same parameters as in Figure~\ref{fig:profil_T} were used.
}
\label{fig:afterexp_T}
\end{figure}

To be complete we now consider the effect on the contrast
of the ballistic expansion of the cloud performed in the experiment.
As in subsection \ref{subsec:tof} we integrate numerically the 
rescaled Gross-Pitaevskii equation, with a classical field $\psi$
including thermal fluctuations. As we see in Figure
\ref{fig:afterexp_T} the rescaling now absorbs to a lower extent the effect
of the ballistic expansion, but the density contrast 
is only weakly changed.

\section{Conclusion}
We have understood why the vortex line can bend in a steady state
cigar-shaped condensate
rotated at frequency $\Omega$
around its long axis $z$. In the Thomas-Fermi regime 
the cigar-shaped condensate can be viewed as a collection of slices
parallel to the $x-y$ plane, each slice corresponding formally to a 2D 
rotating condensate.
For each 2D condensate one defines as usual the critical rotation frequency
$\Omega_{\rm c}^{\rm 2D}$ above which it is energetically more favorable
to have the vortex core at the trap center rather than at infinity.
As the 3D condensate density is inhomogeneous along $z$ axis due to 
the harmonic confinement the local $\Omega_{\rm c}^{\rm 2D}$ is minimal
in the plane $z=0$ and maximal at the end points of the cigar.

The vortex line then uses the following strategy to minimize its energy:
it follows the rotation axis $z$ in the elevation interval
where the trap rotation frequency is larger than the local 
$\Omega_{\rm c}^{\rm 2D}$'s,
and moves away from the rotation axis to infinity where $\Omega$ becomes
smaller than the local $\Omega_{\rm c}^{\rm 2D}$.

This leads to the analytical prediction that the minimal rotation frequency
$\Omega_1$ required to stabilize the bent vortex in a cigar-shaped condensate
is equal to $\Omega_{\rm c}^{\rm 2D}(z=0)$, that is the 2D critical rotation
frequency corresponding to the central slice $z=0$. 

Another analytical prediction is that, for $\Omega>\Omega_1$,
the energy presents a saddle point,  in addition to the previously 
discussed minimum. This saddle point is a stationary configuration
of the vortex line in the rotating frame, with an angular momentum
smaller than the one of the energy minimum. The vortex line in the
saddle point has also a different position with respect
to $z$ axis than in the minimum energy configuration:
the vortex line, rather than being on axis,
is shifted away from the rotation axis by an amount
of the order of the transverse Thomas-Fermi radius $R_\perp$, 
and it has a much smaller length along $z$, also on the
order of $R_\perp$. This description matches very recent experimental
observations of bent vortices at ENS \cite{vu_a_ENS}.

To test the analytical predictions for the energy minimum
we have performed a full numerical minimization of the Gross-Pitaevskii
energy functional with the efficient conjugate gradient method for typical
parameters of the ENS experiment.
We have constructed in this way a phase diagram giving the existence
domain of a single vortex configuration as a local minimum of energy,
as function of the trapping potential aspect ratio and of the
rotation frequency. 
The numerics confirm the analytical prediction for $\Omega_1$,
and give the rotation frequency $\Omega_2$
above which surface modes of the condensate are destabilized and
several vortices enter the condensate. 

We have also studied the effect of thermal fluctuations of an otherwise
straight vortex line in a cigar-shaped condensate. 
We find values of the mean density contrast of the vortex hole
close to the experimental results of \cite{ens}.
Using the Glauber P representation of the density
operator of the gas for the field of Bogoliubov quasi-particles,
we can predict what happens in a single realization of the experiment:
the vortex line experiences some bending, due to the thermal population
of low energy kelvon modes localized close to the end points of the
condensate.  The 1D nature of the kelvon modes 
leads to a remarkable non Gaussian character of the vortex line fluctuations.

\acknowledgments
We acknowledge useful discussions with Bernard Bernu, Alice Sinatra,
Gora Shlyapnikov, Subhasis Sinha,
Amandine Aftalion, Tristan Rivi{\`e}re, Jean Dalibard and the members
of his group.
This work was supported by the  BEC2000+ Programme of ESF. 
M.M. would like to thank the \'Ecole normale 
sup{\'e}rieure in Paris for hospitality.
Laboratoire de Physique Th\'eorique des Liquides is the Unit\'e Mixte
de Recherche 7600 of Centre National de la Recherche Scientifique.
Laboratoire Kastler Brossel is a research unit of 
\'Ecole normale sup\'erieure and of Universit\'e Pierre et Marie
Curie, associated to Centre National de la Recherche Scientifique.

\appendix
\section{Simple derivation of a vortex line energy functional}
\label{appendix}

We present here a derivation of an approximate energy 
functional for a single vortex line in a Thomas-Fermi condensate subject to a
harmonic potential rotating around one of its eigenaxes.
Such a derivation is available in the literature \cite{AftaRiv}
but it is rather complex as it directly includes the effect of curvature 
of the vortex line in the energy functional, and it applies for an arbitrary
aspect ratio of the condensate.
The derivation that we propose has the advantage of simplicity as (i)
it is specialized to the case of a very elongated condensate along the 
rotation axis $z$ in a cylindrically symmetric trap, 
that is the atomic oscillation frequency $\omega_z$
along $z$ is much smaller than the oscillation frequency 
$\omega_\perp$ in the transverse $xy$ plane, and (ii) it immediately supposes that
the explicit dependence of the energy functional on
the curvature of the steady state bent vortex line
can be neglected.
Point (ii) allows to
assume locally that the vortex line is straight, which greatly simplifies
the derivation of the energy functional. 
By comparison with the more general derivation of \cite{AftaRiv}
we have checked the validity of point (ii) \cite{verif_curv}.

Furthermore we assume in our derivation that the rotation frequency
$\Omega$ is of the order of
\begin{equation}
\Omega \sim \frac{\hbar\omega_\perp^2}{\mu}
\label{eq:range}
\end{equation}
where $\omega_\perp$ is the oscillation frequency of the atoms in the
$xy$ plane and $\mu$ is the chemical potential. It is indeed in this range
of rotation frequencies that the single vortex configuration is first stabilized,
and this also greatly simplifies the derivation of the energy functional.
One then finds an energy difference between the vortex free configuration (of energy $E_0$)
and a configuration with a vortex of the order of
\begin{equation}
E - E_0 \sim \frac{\hbar^2\omega_\perp^2}{\mu}.
\label{eq:range_ener}
\end{equation}

The first step in the derivation of the approximate energy functional
is to rewrite the full Gross-Pitaevskii energy functional
(\ref{eq:ener}) in terms of the modulus and the phase of the 
condensate wavefunction $\phi(\vec{r}\,)$:
\begin{equation}
\phi(\vec{r}\,) = n^{1/2}(\vec{r}\,) e^{iS(\vec{r}\,)}
\label{eq:mod_phase}
\end{equation}
where $n$ is the probability density normalized to unity.
At this stage it is convenient to introduce the so-called velocity field
of the condensate:
\begin{equation}
\vec{v}(\vec{r}\,) = \frac{\hbar}{m}\vec{\mbox{grad}}\, S(\vec{r}\,).
\end{equation}
Note that $\vec{v}$  corresponds to the velocity field in the lab frame,
the velocity field in the rotating frame being $\vec{v}-\vec{\Omega}
\land \vec{r}$. Inserting (\ref{eq:mod_phase}) into the Gross-Pitaevskii energy functional
(\ref{eq:ener}) leads to an expression that we split for convenience in two contributions,
one linked to the modulus and the other one linked to the phase:
\begin{eqnarray}
E &=& E_{\rm mod} + E_ {\rm phase} \\
E_{\rm mod} &=& \int d^3\vec{r}\, \left[\frac{\hbar^2}{2m}(\vec{\mbox{grad}}\sqrt{n})^2
+U n + \frac{1}{2}N g n^2\right] 
\label{eq:emod}\\
E_{\rm phase} &=& \int d^3\vec{r}\, n\left[\frac{1}{2}m\vec{v}\,^2-\vec{\Omega}\cdot
(\vec{r}\land m\vec{v})\right].
\label{eq:ephase}
\end{eqnarray}
As already mentioned, the trapping potential $U$ is axi-symmetric with respect to $z$:
\begin{equation}
U = \frac{1}{2} m\left[\omega_\perp^2 (x^2+y^2) +\omega_z^2 z^2\right].
\end{equation}
We also give the conditions on $n$ and $\vec{v}$
ensuring that $\phi$ is a local extremum of the Gross-Pitaevskii
energy functional:
\begin{eqnarray}
0 &=& \mbox{div}\left[n(\vec{v}-\vec{\Omega}\land\vec{r}\,)\right] 
\label{eq:sta1} \\
\mu &=& \frac{1}{2} m\vec{v}\,^2 + U + N g n 
-\frac{\hbar^2}{2m}\frac{\Delta\sqrt{n}}{\sqrt{n}}
-m\vec{\Omega}\land\vec{r}\cdot\vec{v} \nonumber\\
&& \label{eq:sta2}
\end{eqnarray}
where $\mu$ is the chemical potential.
These conditions are simply the time independent Gross-Pitaevskii equation 
written in the modulus-phase representation.

\subsection{Energy from the modulus}\label{subsec:emod}
In the present Thomas-Fermi regime 
the contribution depending only on the modulus can be evaluated along the lines
of our previous work \cite{castindum}. One splits the density in a 
slowly varying envelope $n_{\rm slow}$ and a function $f^2$ representing the density hole
due to the vortex core:
\begin{equation}
n(\vec{r}\,) = n_{\rm slow}(\vec{r}\,)  f^2(\vec{r}\,).
\label{eq:split_mod}
\end{equation}
The envelope $n_{\rm slow}$ varies at the scale of the transverse Thomas-Fermi
radius of the condensate $R_\perp=(2\mu/m\omega_\perp^2)^{1/2}$ whereas
$f$ varies at the scale of the diameter of a vortex core, that
is the healing length $\xi=(\hbar^2/m\mu)^{1/2}$ \cite{convenience}.

We now determine the function $f$ from the requirement that it
deviates significantly from unity only at a distance
at most a few $\xi$'s from the vortex core. As such a length scale we neglect the
spatial variation of $U$ and of $n_{\rm slow}$;  we further check that in the 
range (\ref{eq:range}) of rotation frequencies,
the rotational velocity term $\vec{\Omega}\land\vec{r}$
is negligible as compared to the vortex velocity field $\vec{v}$, which allows to neglect
the $\vec{\Omega}\land\vec{r}$ terms close to the vortex core both in
(\ref{eq:sta1}) and (\ref{eq:sta2}):
\begin{equation}
\frac{|\vec{\Omega}\land\vec{r}\,|}{v} \sim \frac{\Omega R_\perp}{\hbar/m\xi}
\sim \frac{\Omega}{\omega_\perp} \sim \frac{\hbar\omega_\perp}{\mu}
\ll 1.
\label{eq:cond_pas_rot}
\end{equation}
Furthermore the minimal radius of curvature of the vortex line is found in the
subsequent calculations to be of the order of $R_\perp$ much larger than $\xi$.
After all these simplifications the function $f$ is found locally to solve
the well-known Gross-Pitaevskii equation for a straight vortex line in an infinite
spatially homogeneous condensate \cite{Donnelly}, this fictitious homogeneous condensate
having a particle density given by $N n_{\rm slow}$ evaluated on the vortex line.
We therefore take for $f$:
\begin{equation}
f(\vec{r}\,) = F(d/\xi_{\rm loc})
\label{eq:for_f}
\end{equation}
where $d$ is the distance of $\vec{r}$ to the vortex line and 
\begin{equation}
\xi_{\rm loc}= \frac{\hbar}{\sqrt{mgNn_{\rm slow}(\vec{r}_0)}}
\label{eq:xi_loc}
\end{equation}
is the local healing length
at the position of the vortex line \cite{convenience}. The function
$F(u)$ does not depend on any physical parameter. It
can be obtained from a numerical solution of the reduced Gross-Pitaevskii equation
for a vortex in a homogeneous condensate 
(see equation (2.84) of \cite{Donnelly}). In
the large $u$ limit, its deviation from unity tends
to zero as $O(1/u^2)$. For moderate values of $u$
it is well approximated by \cite{castindum}
\begin{equation}
F(u) \simeq \tanh(0.7687 u).
\label{eq:th}
\end{equation}

The slowly varying envelope $n_{\rm slow}$ is obtained from (\ref{eq:sta2}) by removing
the short range $mv^2/2$ term already included in $f$ and by neglecting the quantum
pressure term $\propto \Delta\sqrt{n}/\sqrt{n}$ in the spirit of the Thomas-Fermi
approximation. We write $n_{\rm slow}$ as $n_{\rm TF} + \delta n$ where 
\begin{equation}
n_{TF}(\vec{r}\,) = \frac{\mu_0 - U(\vec{r}\,)}{Ng}
\end{equation}
is the usual Thomas-Fermi approximation for the probability density 
in the absence of vortex and 
\begin{equation}
N g \delta n =\delta \mu + \vec{\Omega}\land\vec{r}\cdot m\vec{v}.
\end{equation}
The density correction $\delta n$ includes the rotational term and 
the deviation $\delta\mu$ of $\mu$ from the vortex free Thomas-Fermi chemical potential $\mu_0$.
Both contributions are of the same small order. Using the estimate $v\sim \hbar/mR_\perp$
we find that the rotational term leads to
\begin{equation}
\frac{\delta n^{\rm rot}}{n_{\rm TF}} \sim \frac{\hbar\Omega}{\mu_0} \sim \left(
\frac{\hbar\omega_\perp}{\mu_0} \right)^2.
\label{eq:term_rot}
\end{equation}
We can also estimate $\delta\mu$ by multiplying (\ref{eq:sta2}) by $n$
and integrating over the whole space: we find a contribution to $\delta\mu$
involving the rotational term,
\begin{equation}
\delta\mu^{\rm rot} = -\Omega \langle L_z\rangle
\end{equation}
of the order of $\hbar\Omega$. A second contribution
comes from the fact that the vortex line digs an empty tube of volume $\sim R_z \xi^2$
in the condensate of volume $\sim R_\perp^2 R_z$, 
where $R_z$ is the Thomas-Fermi radius along $z$, so that $\mu$
has to differ from $\mu_0$ by a relative amount 
\begin{equation}
\frac{\delta \mu^{\rm norm}}{\mu_0} \sim \frac{R_z \xi^2}{R_\perp^2 R_z} 
\sim \left(\frac{\hbar\omega_\perp}{\mu_0}\right)^2
\end{equation}
to ensure that $n$ is normalized to unity.

We now proceed with the calculation of $E_{\rm mod}$, inserting the ansatz
(\ref{eq:split_mod}) in (\ref{eq:emod}). We calculate first the harmonic plus interaction
potential energy part of $E_{\rm mod}$, then the kinetic energy part of $E_{\rm mod}$.

\begin{figure}
\centerline{\includegraphics[width=8cm,clip=]{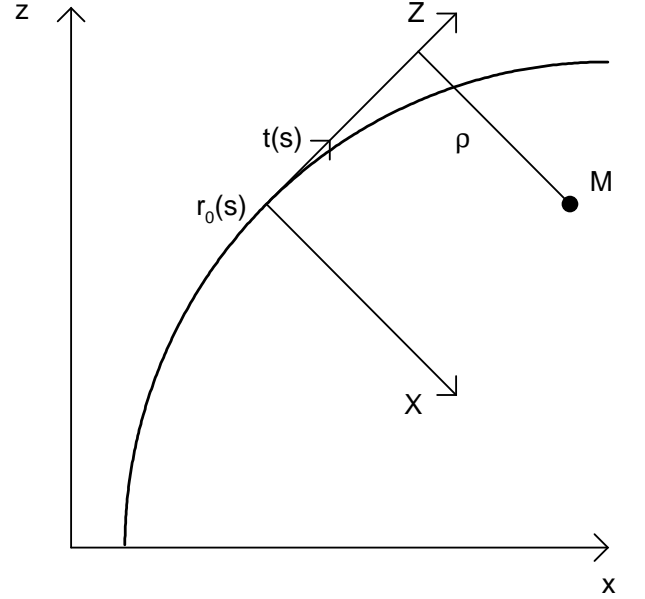}} 
\caption{Local frame with Cartesian coordinates
$X,Y,Z$ around a point $\vec{r}_0(s)$ of the locally straight
vortex line. Local axis $Z$ is tangent to 
the vortex line and has the orientation of the local vorticity,
so that the unit vector along $Z$ coincides with the vector $\vec{t}(s)$
tangent to the vortex line defined in the text. $X$, $Y$ 
are arbitrary Cartesian coordinates in the plane orthogonal to $Z$. One then defines
the corresponding cylindrical coordinates $\rho,\theta,Z$ of an arbitrary
point $M$, {\it e.g.} $\rho$ is the distance of $M$ to the vortex line.
\label{fig:xyz}}
\end{figure}

In the harmonic plus interaction potential energy terms, 
we use the identity $f^2=(f^2-1)+1$ and we collect the terms in
powers of $f^2-1$:
\begin{eqnarray}
E_{\rm pot} &=& E_{\rm pot}^{(0)} + E_{\rm pot}^{(1)} + E_{\rm pot}^{(2)} \\
E_{\rm pot}^{(0)} &=& \int d^3\vec{r}\, 
\left(\frac{N g}{2} n_{\rm slow}^2 + U n_{\rm slow}\right)  \\
E_{\rm pot}^{(1)} &=& \int d^3\vec{r}\, (Ng n_{\rm slow} + U) n_{\rm slow} (f^2-1) \\
E_{\rm pot}^{(2)} &=& \int d^3\vec{r}\, \frac{N g}{2} n_{\rm slow}^2(f^2-1)^2.
\end{eqnarray}
The zeroth degree term is of the order of magnitude of $\mu$ so that one has
to include the deviation of $n_{\rm slow}$ from $n_{\rm TF}$ to first order 
in order to get the leading term (\ref{eq:range_ener}) in the vortex energy. Using
the fact that $Ngn_{\rm TF}+U=\mu_0$ we obtain
\begin{equation}
E_{\rm pot}^{(0)} = E_{\rm TF} + \mu_0 \int d^3\vec{r}\, \delta n + o(\hbar^2\omega_\perp^2/\mu),
\end{equation}
where we have introduced the Thomas-Fermi approximation to the energy $E_0$ of the vortex 
free configuration:
\begin{equation}
E_0 \simeq E_{\rm TF} =  \int d^3\vec{r}\, \left(
\frac{N g}{2} n_{\rm TF}^2 + U n_{\rm TF}\right).
\end{equation}
The first degree term $E_{\rm pot}^{(1)}$  and the second degree term $E_{\rm pot}^{(2)}$
are of the order of $\hbar^2\omega_\perp^2/\mu_0$ as $|f^2-1|$ is close to unity in a cylinder
of volume $R_z \xi^2$ and is negligible outside. We can therefore approximate $n_{\rm slow}$
by $n_{\rm TF}$ in $E_{\rm pot}^{(1)}$ and $E_{\rm pot}^{(2)}$.
We then get for $E_{\rm pot}^{(1)}$:
\begin{eqnarray}
E_{\rm pot}^{(1)} &\simeq& \mu_0 \int d^3\vec{r}\, n_{\rm TF} (f^2-1) \nonumber \\
&=& 
-\mu_0 \int d^3\vec{r}\,  \delta n f^2 \simeq -\mu_0 \int d^3\vec{r}\,  \delta n
\end{eqnarray}
where we have used the normalization of $n_{\rm TF}$ and 
of $n_{\rm slow} f^2=(n_{\rm TF}+\delta n)f^2$ 
to unity. We thus see that at the present order of the calculation $E_{\rm pot}^{(1)}$
compensates the term linear in $\delta n$ in $E_{\rm pot}^{(0)}$. We finally get:
\begin{equation}
E_{\rm pot} - E_{\rm TF} \simeq \int d^3\vec{r}\, \frac{N g}{2} n_{\rm TF}^2(f^2-1)^2.
\end{equation}
In this integral we introduce a local system of cylindrical coordinates $(\rho,\theta,Z)$ 
with a vertical axis $Z$ tangent to the vortex line,
see Figure \ref{fig:xyz}, the first coordinate $\rho$ being the distance to the vortex line.
We then approximate $n_{\rm TF}$ by its local value on the vortex line. The angular integral 
over $\theta$ gives a factor $2\pi$. The function $f^2-1$ depends on the distance $\rho$
to the vortex line, see equation
(\ref{eq:for_f}), and the corresponding integral over $\rho$ can be extended to infinity
as $(f^2-1)^2$ tends rapidly to zero far from the vortex line.
We parametrize the vortex line by its curvilinear abscissa $s$. We then
realize that in the integral over $Z$ we can make the reinterpretation $dZ=ds$ so that
\begin{eqnarray}
&&E_{\rm pot} - E_{\rm TF} \simeq \\
&&\int ds\, \frac{N g}{2} n_{\rm TF}^2(\vec{r}_0(s))
\int_0^{+\infty}\!\!\!2\pi\rho d\rho\,\left[F^2(\rho/\xi_{\rm loc}(s))-1\right]^2
\nonumber
\end{eqnarray}
where $\vec{r}_0(s)$ is the position of the vortex line at abscissa $s$.
Finally rescaling $\rho$ by $\xi_{\rm loc}$ in the integral over $\rho$ and
replacing $n_{\rm slow}$ by $n_{\rm TF}$ in (\ref{eq:xi_loc}) leads to
\begin{equation}
E_{\rm pot} - E_{\rm TF} \simeq A_0\int ds\, \frac{\pi\hbar^2}{m} n_{\rm TF}(\vec{r}_0(s))
\end{equation}
where the constant factor $A_0$ is
\begin{equation}
A_0=\int_0^{+\infty}u \, du\,\left[F(u)^2-1\right]^2.
\end{equation}

In the kinetic energy term of $E_{\rm mod}$,
we neglect the spatial derivative of the slowly varying envelope
$n_{\rm slow}$ and of the local healing length $\xi$,
as they both vary on a length scale $R_\perp$. We calculate the gradient of $f$ in the local system of
cylindrical coordinates $(\rho,\theta,Z)$ of Figure \ref{fig:xyz}:
\begin{equation}
(\vec{\mbox{grad}}f)^2 = \frac{1}{\xi_{\rm loc}^2} F^{'2}(\rho/\xi_{\rm loc}).
\end{equation}
This function vanishes at $\rho\gg \xi_{\rm loc}$ from the vortex line
so that we can replace $n_{\rm slow}$
by its value on the vortex line and extend the integral over $\rho$ to infinity. The integral
over the angle $\theta$ gives a factor $2\pi$. As in the previous paragraph we use
$dZ=ds$ where $s$ is the curvilinear abscissa on the vortex line,
we rescale $\rho$ by $\xi_{\rm loc}$ in the integral over $\rho$ and we replace
$n_{\rm slow}$ by the vortex free Thomas-Fermi expression $n_{\rm TF}$.
We then obtain for the kinetic energy part of $E_{\rm mod}$:
\begin{equation}
E_{\rm mod}^{\rm kin} \simeq A_1 \int ds \frac{\pi\hbar^2}{m}n_{\rm TF}(\vec{r}_0(s))
\end{equation}
where the constant factor $A_1$ is  given by
\begin{equation}
A_1 = \int_0^{+\infty} u du\, F^{'2}(u).
\end{equation}

To summarize the vortex energy stored in the modulus of the condensate wavefunction
is approximated by
\begin{equation}
E_{\rm mod} - E_{\rm TF}\simeq (A_0+A_1) \int ds \frac{\pi\hbar^2}{m}n_{\rm TF}(\vec{r}_0(s)).
\end{equation}
We estimate $E_{\rm mod} - E_{\rm TF}$ to be indeed
of the order of magnitude of $\hbar^2\omega_\perp^2/\mu_0$, as expected
from (\ref{eq:range_ener}),
by taking a length $R_z$ for the part of the vortex line inside the condensate and a 
typical value for the Thomas-Fermi envelope $n_{\rm TF} \sim 1/R_z R_\perp^2$.

\subsection{Energy involving the velocity field}
The energy term $E_{\rm phase}$ is quite difficult to evaluate in its present form 
(\ref{eq:ephase}): since the velocity field tends to zero slowly away from the
vortex line (as $\hbar/md$ where $d$ is the distance to the vortex line), all the parts
of the Thomas-Fermi volume of the condensate give approximately the same contribution.
A direct 3D integration cannot be performed as the velocity field far from the
vortex line is not known explicitly \cite{en_principe}, with the notable exception
of a straight vortex line on axis $z$.

The trick that we borrow from \cite{AftaRiv} is to transform the volume
integral (\ref{eq:ephase}) in a line integral, by \lq integrating by part'
and using the fact that $\vec{\mbox{curl}}\,\vec{v}$ is known exactly:
\begin{equation}
\vec{\mbox{curl}}\,\vec{v}(\vec{r}\,) = \frac{2\pi\hbar}{m}\int ds\, \vec{t}(s) \,
\delta(\vec{r}-\vec{r}_0(s))
\label{eq:rot_connu}
\end{equation}
where $s$ is the curvilinear abscissa along the vortex line
and $\vec{t}$ is the unit vector tangent to the vortex line and oriented in the direction
of the local vorticity ($\vec{t}$ has a positive component along $z$ for a positive rotation
frequency $\Omega$). Physically one can use an electromagnetic analogy:
$\vec{\mbox{curl}}\,\vec{v}$ corresponds to a linear current, that is
a perfectly filiform charge current,
of intensity $2\pi\hbar/m$ circulating in the vortex line.

The \lq integration by part' to be performed relies on the following vectorial identity
\begin{equation}
\mbox{div}(\vec{A}\land\vec{B}) = \vec{B}\cdot\vec{\mbox{curl}}\,\vec{A}
-\vec{A}\cdot\vec{\mbox{curl}}\,\vec{B}.
\end{equation}
Integrating this identity over whole space and using Ostrogradski's formula we arrive
at the desired identity
\begin{equation}
\int\, d^3\vec{r}\, \vec{A}\cdot\vec{\mbox{curl}}\,\vec{B} =
\int\, d^3\vec{r}\, \vec{B}\cdot\vec{\mbox{curl}}\,\vec{A} 
\label{eq:ipp}
\end{equation}
where we have assumed that the total flux of the vector $\vec{A}\land\vec{B}$ vanishes
through a surface at infinity.

We apply (\ref{eq:ipp}) first to transform the contribution $E_{\rm phase}^{\rm kin}$
of $\vec{v}\,^2$ to (\ref{eq:ephase}):
\begin{equation}
E_{\rm phase}^{\rm kin} = \int d^3\vec{r}\, \frac{1}{2}mn\vec{v}\,^2.
\end{equation}
 We have to choose $\vec{A}=\vec{v}$ so that $\vec{\mbox{curl}}\,\vec{v}$
appears in the left hand side of (\ref{eq:ipp}). 
We then set \cite{different}:
\begin{equation}
\vec{\mbox{curl}}\,\vec{B}\simeq n\vec{v}.
\end{equation}
This is an approximation as $n\vec{v}$ does not have strictly speaking a vanishing
divergence. The vector which has an exactly vanishing divergence is the
probability current in the rotating frame,  $n(\vec{v}-\vec{\Omega}\land\vec{r}\,)$,
see (\ref{eq:sta1}). Fortunately we can repeat the reasoning of subsection
(\ref{subsec:emod}). At a distance of up to a few healing lengths $\xi$
from the vortex core, the rotation term $\vec{\Omega}\land\vec{r}$ is negligible
as compared to the velocity field $\vec{v}$, see (\ref{eq:cond_pas_rot}) 
so that (\ref{eq:sta1}) reduces to $\mbox{div}(n\vec{v}\,)\simeq 0$. At a distance 
much larger than $\xi$ from the vortex core the density $n$ can be approximated 
by the vortex free
Thomas-Fermi density $n_{\rm TF}$ so that (\ref{eq:sta1}) reduces to
\begin{equation}
\mbox{div}\left[n_{\rm TF}(\vec{v}-\vec{\Omega}\land\vec{r}\,)\right] \simeq 0.
\label{eq:interm}
\end{equation}
As the Thomas-Fermi density $n_{\rm TF}$ is rotationally symmetric with respect to 
the rotation axis $z$, we have \cite{pourquoi}
\begin{equation}
\mbox{div}\left[n_{\rm TF}\vec{\Omega}\land\vec{r}\right] = 0
\label{eq:div_nulle}
\end{equation}
so that (\ref{eq:interm}) reduces to $\mbox{div}(n\vec{v}\,)\simeq 0$.
This finally leads to the desired line integral reformulation
for $E_{\rm phase}^{\rm kin}$:
\begin{equation}
E_{\rm phase}^{\rm kin} \simeq \pi\hbar \int ds\, \vec{B}\,(\vec{r}_0(s)) \cdot\vec{t}\,(s).
\end{equation}

The calculation of $\vec{B}$ remains a challenge. Formally $\vec{B}$ can be considered
as the static magnetic field created by a current proportional to $n\vec{v}$
so that we have the Biot and Savart formula \cite{Jackson}:
\begin{equation}
\vec{B}\,(\vec{r}\,) = \frac{1}{4\pi} \int d^3\vec{r}\,'\, 
\frac{n\vec{v}\,(\vec{r}\,')\land(\vec{r}-\vec{r}\,')}{|\vec{r}-\vec{r}\,'|^3}
\label{eq:Biot}
\end{equation}
but this requires in principle the knowledge of the velocity field $\vec{v}$
everywhere \cite{terme_de_bord}. In practice the problem is simplified by the fact that 
one needs to know $\vec{B}$ on the vortex line only, $\vec{r}=\vec{r}_0(s)$ for any fixed $s$,
and by the fact that the integrand
(\ref{eq:Biot}) tends rapidly to zero for increasing $|\vec{r}_0(s)-\vec{r}\,'|$ so that 
it is sufficient to know the velocity field $\vec{v}$ close to the vortex line.
As mentioned in the beginning of this appendix
we assume that the vortex line is locally straight 
around the point $\vec{r}_0(s)$. We then approximate $\vec{v}$ by the velocity field
of a straight vortex in a homogeneous medium. In the local system of cylindrical
coordinates $(\rho,\theta,Z)$ defined in Figure \ref{fig:xyz} we thus write
\begin{equation}
\vec{v}(\vec{r}\,') \simeq \frac{\hbar}{m} \frac{\vec{e}_\theta}{\rho}.
\end{equation}
In this local frame the vector $\vec{r}\,'-\vec{r}_0(s)$ is equal to
$\rho \vec{e}_\rho + Z \vec{e}_Z$ and the unit vector $\vec{e}_Z$
tangent to the vortex line actually coincides with $\vec{t}\,(s)$ so that one has
\begin{equation}
\vec{t}\,(s) \cdot \left[ \vec{v}(\vec{r}\,') \land (\vec{r}_0(s)-\vec{r}\,') \right]
\simeq \frac{\hbar}{m}.
\end{equation}
This leads to the rather explicit expression
\begin{equation}
E_{\rm phase}^{\rm kin} \sim \frac{\hbar^2}{4m} \int\, ds\,
\int\,d^3\vec{r}\,' \frac{n(\vec{r}\,')}{|\vec{r}_0(s)-\vec{r}\,'|^3}.
\label{eq:presque}
\end{equation}

To calculate (\ref{eq:presque}) we write $n=n_{\rm slow} f^2 \simeq  n_{\rm TF}(f^2-1)+
n_{\rm TF}$ as in subsection
\ref{subsec:emod}. This leads to a splitting of 
$E_{\rm phase}^{\rm kin}$ in two pieces. 

The piece involving $f^2-1$ is re-expressed in terms
of the local cylindrical coordinates $(\rho,\theta,Z)$ of the local $XYZ$ frame
of Figure \ref{fig:xyz}:
\begin{eqnarray}
E_{\rm phase}^{\rm kin}(\mbox{I}) &\equiv& \frac{\hbar^2}{4m}\int ds
\int d^3\vec{r}\,' \frac{n_{\rm TF}(\vec{r}\,')[f^2(\vec{r}\,')-1]}
{\left(|\vec{r}_0(s)-\vec{r}\,'|^2+\epsilon^2\right)^{3/2}} \nonumber\\
&=&\frac{\hbar^2}{4m} 
\int ds \int d^3\vec{R} \frac{n_{\rm TF}(\vec{r}\,')[F^2(\rho/\xi_{\rm loc}(s))-1]}
{\left(\rho^2+Z^2+\epsilon^2\right)^{3/2}}\nonumber
\end{eqnarray}
where we introduced an arbitrarily small $\epsilon$ in the denominator to prevent a divergence
of the integral. The integrand as a function of $Z$ is tending to zero as soon as 
$|Z|$ exceeds a few times $\rho$. As $\rho$ is limited to a few times $\xi_{\rm loc}(s)$
by the function $F^2-1$, the integrand becomes negligible as soon as $|Z|$ exceeds
a few times $\xi_{\rm loc}(s)$, which allows to approximate the slowly varying
Thomas-Fermi envelope $n_{\rm TF}$ by its value on the vortex line and to extend the
integration over $Z$ to infinity. Using 
\begin{equation}
\int_{-\infty}^{+\infty} dZ\, \frac{1}{(Z^2+\rho^2+\epsilon^2)^{3/2}} =
\frac{2}{\rho^2+\epsilon^2}
\label{eq:integrale}
\end{equation}
and extending in the resulting integral the integration over $\rho$ to $\infty$
we obtain the result
\begin{equation}
E_{\rm phase}^{\rm kin}(\mbox{I}) \simeq \frac{\pi\hbar^2}{m} \int\! ds\, n_{\rm TF}(\vec{r}_0(s))
\int_0^{+\infty}\!\!\!\rho d\rho\,\frac{F^2(\rho/\xi_{\rm loc}(s))-1}{\rho^2+\epsilon^2}.
\end{equation}
A more explicit form will be given in the next subsection.

The remaining piece of $E_{\rm phase}^{\rm kin}$ involves the Thomas-Fermi
envelope only:
\begin{equation}
E_{\rm phase}^{\rm kin}(\mbox{II}) = \frac{\hbar^2}{4m}\int\, ds\, \int\,d^3\vec{r}\,' 
\frac{n_{\rm TF}(\vec{r}\,')}{\left(|\vec{r}_0(s)-\vec{r}\,'|^2+\epsilon^2\right)^{3/2}}.
\label{eq:II}
\end{equation}
We simplify this expression by taking advantage of the cigar shaped
nature of the condensate. We first integrate (\ref{eq:presque}) over $z'$.
As in the previous paragraph we use the fact that the integral 
\begin{equation}
\int_{z_0(s)-\zeta}^{z_0(s)+\zeta} dz'\, \frac{1}{|\vec{r}_0(s)-\vec{r}\,'|^3}
\end{equation}
converges to its $\zeta=+\infty$ value as soon as $\zeta$ exceeds a few times $|\vec{r}_{0\perp}(s)
-\vec{r}\,'_\perp|$, where $\vec{r}_\perp$ is the projection of the vector $\vec{r}$
in the $xy$ plane. The range of $|\vec{r}\,_\perp|$ is limited to the transverse Thomas-Fermi
radius of the condensate by the presence of $n_{\rm TF}(\vec{r}\,')$ in the integrand.
The range of $|\vec{r}_{0\perp}(s)|$ exceeds $R_\perp$ for a bent vortex line
as the vortex line gets out of the Thomas-Fermi profile of the condensate; 
however the contribution to the energy
of the vortex line segments at a distance exceeding a few  $R_\perp$'s becomes much
smaller than (\ref{eq:range}) and is hence negligible, see the appendix \ref{appen:loin}.
We can therefore assume that $|\vec{r}_{0\perp}(s) -\vec{r}\,'_\perp|$ is at most a few times
$R_\perp$ in (\ref{eq:II}) so that the integral over $z'$ converges over a distance
of $R_{\perp}$. At such a length scale along the rotation axis $z$,
$n_{\rm TF}(x',y',z')$ is almost constant for a cigar shaped condensate
and can be approximated by $n_{\rm TF}(x',y',z_0(s))$. We then extend the integration
over $z'$ to infinity and we use (\ref{eq:integrale}) to get
\begin{equation}
E_{\rm phase}^{\rm kin}(\mbox{II}) \simeq \frac{\hbar^2}{2m}\int ds \int dx'\,dy'\,
\frac{n_{\rm TF}(x',y',z_0(s))}{|\vec{r}_{0\perp}(s)-\vec{r}\,'|^2+\epsilon^2}.
\end{equation}
The resulting integral over $x',y'$ can be calculated exactly and was already
encountered in the 2D calculation of \cite{castindum}. We will give the result in the
next subsection.

Finally we apply the \lq integration by part' technique (\ref{eq:ipp}) 
to the last term of $E_{\rm phase}$, the rotational energy term:
\begin{equation}
E_{\rm phase}^{\rm rot} \equiv \int d^3\vec{r}\, n\left[-\vec{\Omega}\cdot
\vec{r}\land m\vec{v}\,\right] = -m \int d^3\vec{r}\, \vec{v}\cdot (n\vec{\Omega}\land\vec{r}\,).
\end{equation}
Note that this term is simply $-\Omega \langle L_z\rangle$, where $\langle L_z\rangle$
is the angular momentum per particle along $z$.
Calculation of this term is considerably simpler than
$E_{\rm phase}^{\rm kin}$.
First one can approximate $n$ by the Thomas-Fermi envelope $n_{\rm TF}$, neglecting in
particular the density hole due to the vortex line \cite{plus_facile}. Then one realizes
that $n_{\rm TF}\vec{\Omega}\land\vec{r}$, having a vanishing divergence, see
(\ref{eq:div_nulle}), can be written as the curl of some vectorial field $\vec{B}$.
One finds inside the Thomas-Fermi condensate \cite{AftaRiv}
\begin{equation}
\vec{B} = \frac{Ng}{2m\omega_\perp^2} n_{\rm TF}^2\vec{\Omega}
\end{equation}
and one takes $\vec{B}=\vec{0}$ out of the Thomas-Fermi condensate.
One then uses (\ref{eq:ipp}) and (\ref{eq:rot_connu}) to obtain
\begin{equation}
E_{\rm phase}^{\rm rot} \simeq  - \frac{\pi\hbar N g \Omega}{m\omega_\perp^2}
\int ds\, \vec{t}\,(s)\cdot\vec{e}_z \,n_{\rm TF}^2(\vec{r}_0(s))
\end{equation}
where $\vec{e}_z$ is the unit vector along $z$. Note that $\vec{t}\,(s)\cdot\vec{e}_z=\cos\alpha(s)$
where $\alpha$ is the angle between the vortex line and the rotation axis $z$, so that
the integration element $ds\, \cos\alpha(s)$ is simply $dz$, the length of the projection
of the vortex line along $z$. From the estimate $n_{\rm TF}\sim 1/R_\perp^2 R_z\sim \mu_0/N g$
and for a vortex length $\sim R_z$ inside the Thomas-Fermi envelope one gets
$E_{\rm phase}^{\rm rot} \sim -\hbar\Omega$, as expected from $\langle L_z\rangle\sim \hbar$
for a single vortex configuration. This is in the energy range (\ref{eq:range_ener})
for a rotation frequency of the order of (\ref{eq:range}).

\subsection{In terms of a 2D energy functional}

We now reexpress the 3D vortex line energy functional that we have
obtained in terms of the energy functional of a 2D condensate with
a vortex.  Physically we associate a 2D fictitious condensate to each horizontal slice 
of the 3D condensate. Each 2D condensate is stored in a isotropic harmonic potential
$m\omega_\perp^2(x^2+y^2)/2$ with a rotation frequency $\Omega$, and
has the same number $N$ of particles as the 3D condensate. 
The fictitious 2D condensate at elevation $z$ has a vortex at a position given
by the intersection of the 3D vortex line with the horizontal plane of elevation $z$.
The 2D condensate has a Thomas-Fermi chemical potential 
\begin{equation}
\mu_{\rm 2D} (z) = \mu_0 - \frac{1}{2} m\omega_z^2 z^2
\end{equation}
where $\mu_0$ is the 3D vortex free Thomas-Fermi chemical potential. 
The Thomas-Fermi radius of the 2D condensate
therefore coincides with the one $R_\perp(z)$
of the 3D condensate at elevation $z$:
\begin{equation}
R_\perp(z) = \left(\frac{2\mu_{\rm 2D}(z)}{m\omega_\perp^2}\right)^{1/2}.
\end{equation}
Using the Thomas-Fermi
value of the chemical potential for a 2D condensate we get the effective 2D coupling
constant of the 2D condensates:
\begin{equation}
g_{\rm 2D}(z) = \frac{\pi\mu_{\rm 2D}^2(z)}{Nm\omega_\perp^2}.
\label{eq:g_2D}
\end{equation}
We can then relate the 2D Thomas-Fermi density of the 2D condensate to the Thomas-Fermi
3D density of the true condensate:
\begin{equation}
\mu_0-U(\vec{r}\,) = g n_{\rm TF}(\vec{r}\,) = g_{\rm 2D}(z) n_{\rm 2D}(\vec{r}_\perp;z).
\end{equation}

\begin{widetext}
Collecting all the energy terms of this section and after lengthy calculations
we produce the result
\begin{equation}
E - E_{\rm TF} \simeq
\int ds\, \frac{g_{\rm 2D}(z_0(s))}{g} 
\left[E_{\rm 2D}^{\Omega=0}(r_{0\perp}(s);z_0(s))
+ \cos(\alpha(s)) E_{\rm 2D}^{\rm rot}(r_{0\perp}(s);z_0(s))\right]
\label{eq:final}
\end{equation}
where $\alpha(s)$ is the angle between the vortex line and the axis $z$.
The first 2D energy functional term contains the kinetic energy and the 
harmonic plus interaction
potential energy. For a vortex core inside the Thomas-Fermi radius of the condensate,
that is for $\tilde{r}_{\perp}\equiv r_{\perp} / R_{\perp}(z) < 1$, it is given by
\begin{equation}
E_{\rm 2D}^{\Omega=0}(r_{\perp};z) = \frac{(\hbar\omega_\perp)^2}{\mu_{\rm 2D}(z)}
\left\{ \frac{1}{2} +\left(1-\tilde{r}_{\perp}^2\right) \left[
C+\log\left[\frac{\mu_{\rm 2D}(z)}{\hbar\omega_\perp}\left(1-\tilde{r}_{\perp}^2\right)\right] \right] \right\}
\label{eq:bout1}
\end{equation}
where the value of the constant $C$ is
\begin{equation}
C = A_0+A_1-1+\frac{1}{2}\log 2 +\int_0^1du\, \frac{F^2(u)}{u}
+\int_1^{+\infty} \frac{F^2(u)-1}{u}.
\end{equation}
\end{widetext}
For $\tilde{r}_{\perp}>1$ the vortex line at elevation $z$ is out of the Thomas-Fermi
condensate and only $E_{\rm phase}^{\rm kin}$ has a non-vanishing contribution
to $E_{\rm 2D}^{\Omega=0}$:
\begin{equation}
E_{\rm 2D}^{\Omega=0}(r_{\perp};z) = \frac{(\hbar\omega_\perp)^2}{2\mu_{\rm 2D}(z)}
\left[1+\left(1-\tilde{r}_{\perp}^2\right)\log\frac{\tilde{r}_{\perp}^2}{\tilde{r}_{\perp}^2-1}\right].
\label{eq:bout2}
\end{equation}
The second energy functional term in (\ref{eq:final}) is the rotational energy;
it is given for $\tilde{r}_\perp<1$ by
\begin{equation}
E_{\rm 2D}^{\rm rot}(r_{\perp};z) \simeq  -\hbar\Omega (1-\tilde{r}_{\perp}^2)^2
\label{eq:ener_rot}
\end{equation}
and it vanishes out of the condensate, that is for $\tilde{r}_\perp>1$.
If one uses the hyperbolic tangent estimate of \cite{castindum} for $F$,
see (\ref{eq:th}), one gets $C\simeq 0.0884$. The 2D energy functional $E_{\rm 2D}^{\Omega=0} +E_{\rm 2D}^{\rm rot}$ 
then coincides exactly with the one of equation (64) in \cite{castindum}.

\subsection{Improving the 2D energy functional}
By a more detailed analysis than in \cite{castindum} of the properties of the 2D energy
functional $E_{\rm 2D}^{\Omega=0}$ we have identified a pathology that lead to problems
in the full 3D energy functional minimization:
the derivative of $E_{\rm 2D}^{\Omega=0}(r_{\perp};z)$ with respect to $\vec{r}_\perp$ 
presents a logarithmic singularity on the border of the Thomas-Fermi condensate
$r_{\perp} = R_\perp(z)$, that is it diverges logarithmically to $+\infty$
in $\tilde{r}_\perp = 1^-$ and to $-\infty$ in $\tilde{r}_\perp = 1^+$. As a consequence
$E_{\rm 2D}^{\Omega=0}(r_{\perp};z)$ has a local minimum inside the Thomas-Fermi condensate,
at a distance of the order of the healing length $\xi_{\rm 2D}=
\hbar/(m\mu_{\rm 2D}(z))^{1/2}$ from the
border, see Figure \ref{fig:log_sing}. This local minimum of energy is an artifact of the 
approximations used in the appendix: it is found by a numerical minimization of the full
Gross-Pitaevskii energy functional with imaginary time propagation, that a vortex is 
going
out of the condensate to infinity in the absence of trap rotation.

This artifact was not relevant in the 2D case, one just had to restrict
to the Thomas-Fermi limit $\mu_{\rm 2D}\gg\hbar\omega_\perp$
with vortex cores inside the Thomas-Fermi condensate (excluding a thin layer
of thickness $\simeq \xi_{\rm 2D}$ around the boundary). In the 3D case 
this artifact cannot easily be avoided: the bent vortex necessarily
tries to cross the boundary of the condensate, where it encounters the logarithmic singularity
and can remain trapped. This problem is also worse in 3D because the 3D vortex line 
can explore the extremities of the cigar shaped condensate, where 
the local 2D chemical potential $\mu_{\rm 2D}(z)$ can be of the order of
$\hbar\omega_\perp$; the local minimum artifact in $E_{\rm 2D}^{\Omega=0}$ then becomes
very pronounced, see Figure \ref{fig:log_sing}b,
which seriously prevents the vortex line from bending and leaving the condensate.

\begin{figure}
\centerline{ \includegraphics[width=7cm,clip=]{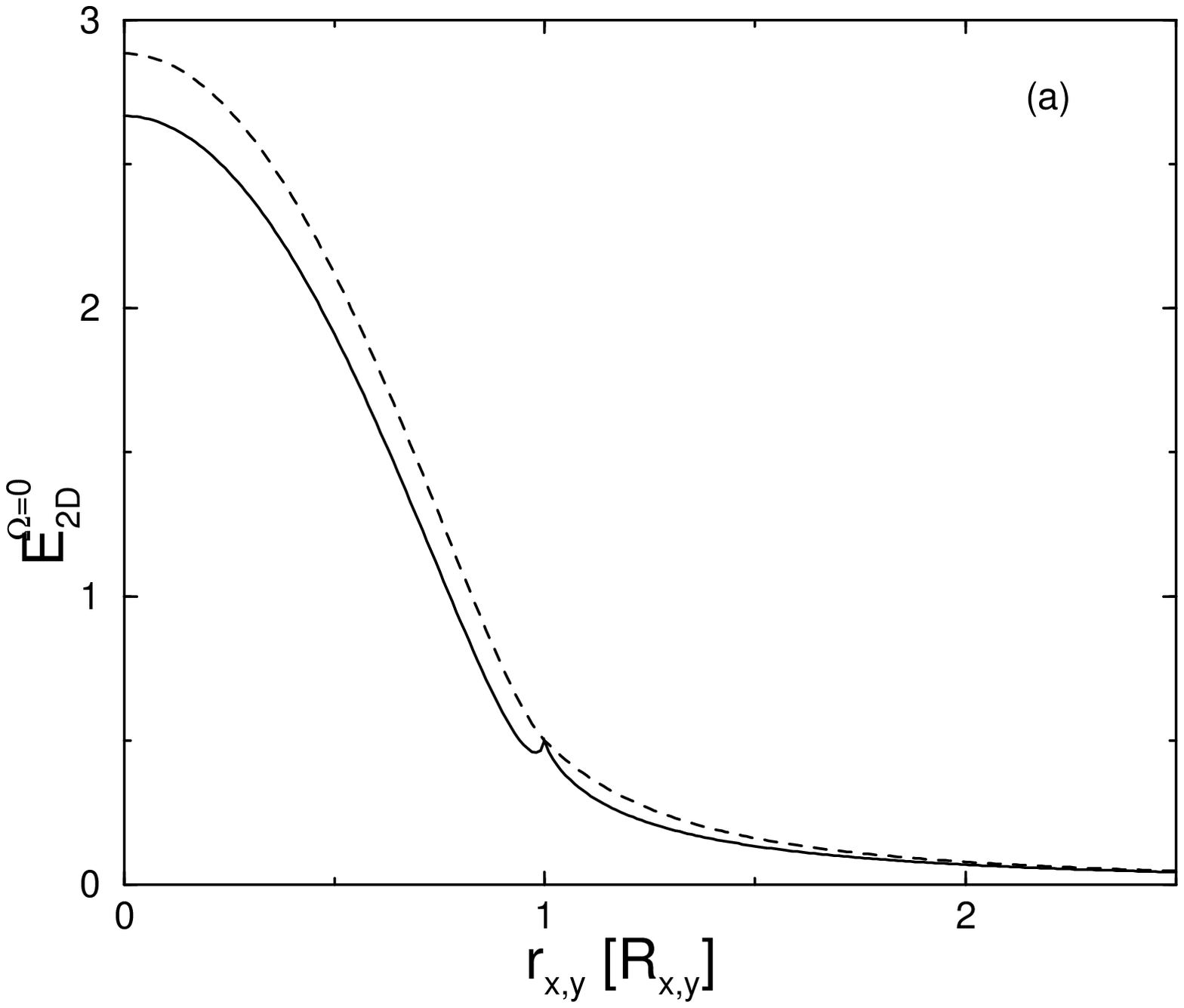}}
\centerline{ \includegraphics[width=7cm,clip=]{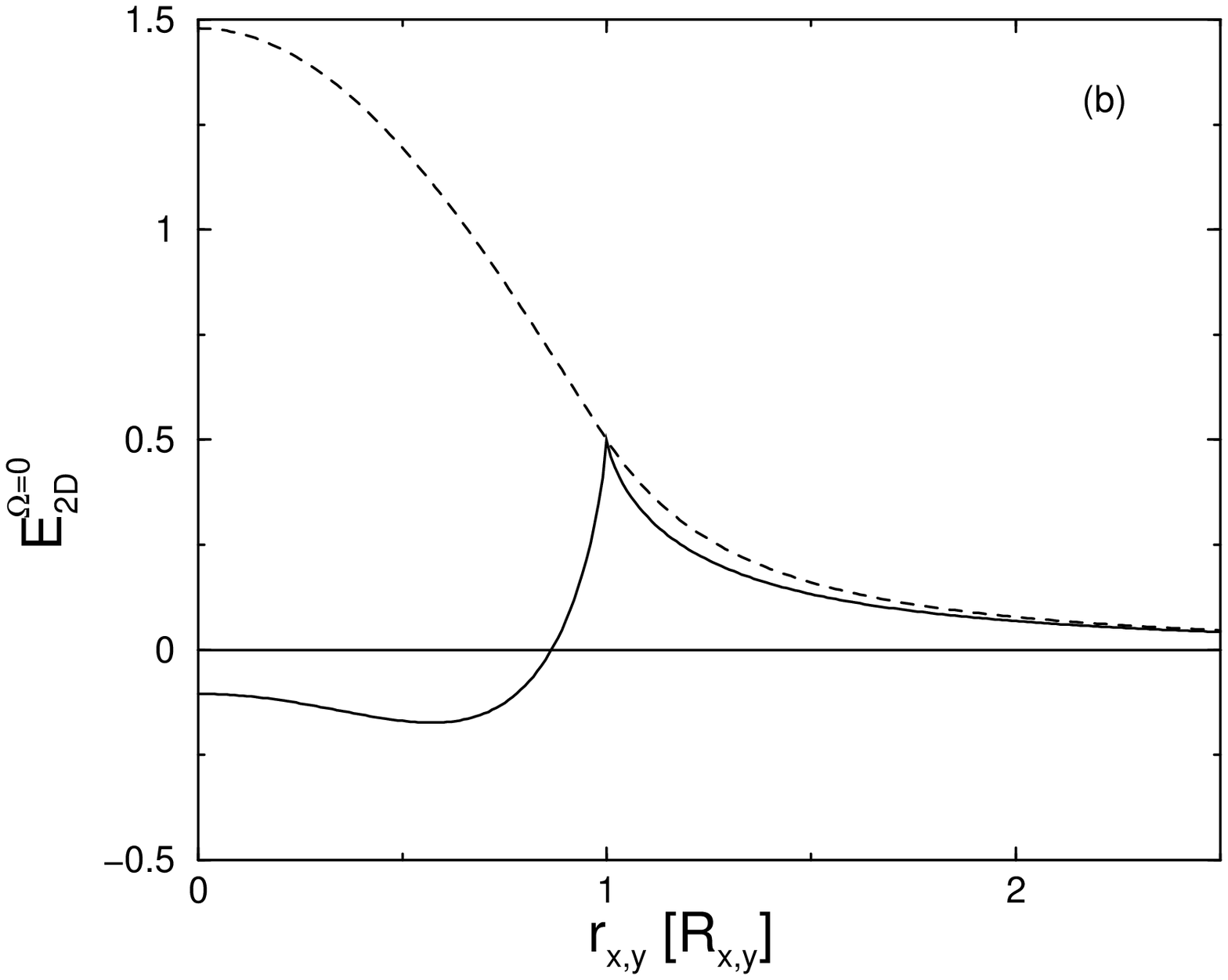}}
\caption{
\label{fig:log_sing}
For a 2D condensate in a non-rotating harmonic trap, energy
of a vortex as function of the vortex
distance $r_\perp$ to the trap center, for a fixed chemical potential
(a) $\mu_{\rm 2D} = 8\hbar\omega_\perp$ and 
(b) $\mu_{\rm 2D} = \hbar\omega_\perp$.
Solid line: the original energy functional of \protect\cite{castindum}, see equations
(\ref{eq:bout1}) and (\ref{eq:bout2}), which presents a spurious local minimum
inside the Thomas-Fermi condensate. Dashed line: the $\eta$-regularized energy
functional heuristically proposed here.}
\end{figure}

Heuristically we perform a modification to the energy functional $E_{\rm 2D}^{\rm \Omega=0}$
eliminating the spurious trapping of the vortex line in the absence of rotation. 
In the region out of the Thomas-Fermi condensate,
$\tilde{r}_\perp > 1$,  we realized that $E_{\rm 2D}^{\rm \Omega=0}$
is well approximated by a small $1/\tilde{r}_\perp$ expansion of the logarithmic term
in (\ref{eq:bout2}) so that we take
\begin{equation}
\bar{E}_{\rm 2D}^{\Omega=0}(r_{\perp};z) = \frac{(\hbar\omega_\perp)^2}{2\mu_{\rm 2D}(z)}
\left(\frac{1}{2\tilde{r}_\perp^2}+\frac{1}{2\tilde{r}_\perp^4}\right).
\label{eq:eta_out}
\end{equation}
In the inner Thomas-Fermi region, $\tilde{r}_\perp < 1$, we
eliminate the logarithmic singularity of the derivative at the inner border
of the condensate by adding a constant term $\eta$ to the argument 
of the logarithm of (\ref{eq:bout1}): 
\begin{widetext}
\begin{equation}
\bar{E}_{\rm 2D}^{\Omega=0}(r_{\perp};z) = \frac{(\hbar\omega_\perp)^2}{\mu_{\rm 2D}(z)}
\left\{ \frac{1}{2} +\left(1-\tilde{r}_{\perp}^2\right) \left[
C+\log\left[\eta+
\frac{\mu_{\rm 2D}(z)}{\hbar\omega_\perp}
\left(1-\tilde{r}_{\perp}^2\right)\right] \right] \right\}.
\label{eq:eta_in}
\end{equation}
\end{widetext}
These definitions lead to a continuous function $\bar{E}_{\rm 2D}^{\Omega=0}$.
The value of $\eta$ is adjusted to further ensure continuity of the derivatives
of $\bar{E}_{\rm 2D}^{\Omega=0}$ in $\tilde{r}_\perp =1$:
\begin{equation}
\eta =  e^{3/4-C} \simeq 1.938.
\end{equation}
As seen on Figure \ref{fig:log_sing} the regularized function 
$\bar{E}_{\rm 2D}^{\Omega=0}$ has no local minimum, even for
a chemical potential smaller than $\hbar\omega_\perp$.

In 2D the physical predictions derived from $\bar{E}_{\rm 2D}^{\Omega=0}$ 
slightly differ in the Thomas-Fermi regime from the ones from
$E_{\rm 2D}^{\Omega=0}$.
For example the critical rotation frequency $\Omega_{\rm c}^{\rm 2D}$
such that the single vortex configuration has the same energy
as the vortex free configuration is, from $\bar{E}_{\rm 2D}^{\Omega=0}$:
\begin{equation}
\bar{\Omega}_{\rm c}^{\rm 2D} = \frac{\hbar\omega_\perp^2}{\mu_{\rm 2D}} 
\log\left[e^{C+1/2}\left(\frac{\mu_{\rm 2D}}{\hbar\omega_\perp}
+\eta\right)
\right]
\label{eq:omega_c_eta}
\end{equation}
whereas the original one in \cite{castindum} is
\begin{equation}
\Omega_{\rm c}^{\rm 2D} = \frac{\hbar\omega_\perp^2}{\mu_{\rm 2D}}
\log\left[e^{C+1/2}\left(\frac{\mu_{\rm 2D}}{\hbar\omega_\perp}
\right) \right].
\end{equation}
In the Thomas-Fermi limit $\bar{\Omega}_{\rm c}^{\rm 2D} $ and $\Omega_{\rm c}^{\rm 2D}$ differ
by a term scaling as $\eta\hbar^3\omega_\perp^3/\mu_{\rm 2D}^2$ which is beyond
accuracy of the energy functional derivation of the present appendix.
The stabilization rotation frequency $\Omega_{\rm stab}^{\rm 2D}$
such that the single vortex core is a local minimum of energy is also modified:
\begin{equation}
\bar{\Omega}_{\rm stab}^{\rm 2D} = \frac{\hbar\omega_\perp^2}{2\mu_{\rm 2D}} 
\log\left[e^{C+1/(1+\eta\hbar\omega_\perp/\mu_{\rm 2D})}
\left(\frac{\mu_{\rm 2D}}{\hbar\omega_\perp}
+\eta\right)
\right]
\label{eq:omega_sta_eta}
\end{equation}
but this varies only to second order in $\eta\hbar\omega_\perp/\mu_{\rm 2D}$.

\section{Energy of the vortex line segments far from the condensate}
\label{appen:loin}

We give here the behaviour of the field $\vec{B}$ defined by (\ref{eq:Biot}) 
far from the Thomas-Fermi condensate, that is at a distance $r$ 
much larger than the Thomas-Fermi radii of the condensate.
This problem is formally equivalent to the calculation in magnetostatics
of the magnetic field $\vec{B}$ very far from a localized distribution of current 
$\vec{j}= n\vec{v}/\mu_0$ where $\mu_0$ is the magnetic permeability of
vacuum. This calculation is performed for example in \cite{Jackson}.
 From the property $\mbox{div} \vec{j}=0$ one finds that the leading term
in the $1/r$ expansion corresponds to a dipolar field created by the magnetic dipole moment
$\vec{M}$ of the current distribution:
\begin{equation}
\vec{B}\,(\vec{r}\,) \simeq \frac{\mu_0}{4\pi r^3} \left[3\vec{u}\,(\vec{u}
\cdot\vec{M}\,) - \vec{M}\,\right]
\end{equation}
where $\vec{u}=\vec{r}/r$ and
the moment $\vec{M}$ is here proportional to the mean angular
momentum per particle in the condensate:
\begin{equation}
\vec{M} \equiv \frac{1}{2} \int d^3\vec{r}\,'\, \vec{r}\,'\land \vec{j}\,(\vec{r}\,')
= \frac{1}{2m\mu_0} \langle\vec{L}\rangle.
\end{equation}
As a consequence $\vec{B}$ tends to zero as $1/r^3$ at infinity, so that
the total flux of $\vec{v}\land\vec{B}$  vanishes at infinity,
as it was assumed in (\ref{eq:ipp}).
When the condensate wavefunction is symmetric under reflection with respect to
the $xy$ plane, as expected for a planar bent vortex contained in the $xz$ plane,
the $x$ and $y$ component of $\langle\vec{L}\rangle$ vanish and we get
\begin{equation}
\vec{B}\,(\vec{r}\,) \simeq \frac{\langle L_z\rangle}{8\pi m}
\frac{3\vec{u}\,(\vec{u}\cdot\vec{e}_z)-\vec{e}_z}{r^3}.
\end{equation}


\begin{thebibliography}{}
\bibitem{ens} 
K.~W.~Madison, F.~Chevy, W.~Wohlleben, J.~Dalibard, 
Phys. Rev. Lett. {\bf 84}, 806 (2000); 
F.~Chevy, K.~W.~Madison, J.~Dalibard, 
Phys. Rev. Lett. {\bf 85}, 2223 (2000).

\bibitem{MIT}
J.R. Abo-Shaeer, C. Raman, J.M. Vogels, and W. Ketterle, Science {\bf
292}, 476 (2001).

\bibitem{Foot}
E. Hodby, G. Hechenblaikner, S. A. Hopkins, O. M. Marag{\`o}, and C. J. Foot,
Phys. Rev. Lett. {\bf 88}, 010405 (2002).

\bibitem{Fetter_review} 
See the recent review by
A.~Fetter, A.~Svidzinsky, J. Phys.: Condens. Matter {\bf 13},  R135-R194 (2001).

\bibitem{Rokhsar}
D. A. Butts and D. S. Rokhsar, Nature {\bf 397}, 327 (1999).

\bibitem{castindum} 
Y.~Castin and R.~Dum, Eur. Phys. J. D {\bf 7}, 399 (1999).

\bibitem{du}
A. Aftalion and  Qiang Du, Phys. Rev. A {\bf 64}, 063603 (2001).

\bibitem{Donnelly} R.J. Donnelly, {\it Quantized vortices in helium II}
(Cambridge, 1991).




\bibitem{garcia} 
J.~J.~Garcia-Ripoll and V.~M.~Perez-Garcia,
preprint {\tt cond-mat/0006368};
Phys. Rev. A {\bf 63}, 041603 (2001);
Phys. Rev. A {\bf 64}, 053611 (2001).

\bibitem{AftaRiv} A. Aftalion, T. Rivi{\`e}re, Phys. Rev. A {\bf 64}, 043611 (2001).

\bibitem{feder}
D.~L.~Feder, A.~A.~Svidzinsky, A.~L.~Fetter and C.~W.~Clark,
Phys. Rev. Lett. {\bf 86}, 564 (2001).

\bibitem{Fetter_dynamics} 
A. Svidzinsky and  A. Fetter, Phys. Rev. A {\bf 62}, 063617 (2000).

\bibitem{numrec} 
 W. H. Press, B. P. Flannery,
S. A. Teukolsky, W. T. Vetterling, {\it Numerical Recipes}, {\S} 10.6,
Cambridge University Press (1986).

\bibitem{dalfovo}
F.~Dalfovo and S.~Stringari, Phys. Rev. A {\bf 63}, 011601(R), (2000).

\bibitem{zambelli} 
F.~Zambelli and S.~Stringari, Phys. Rev. Lett. {\bf 81}, 1754 (1998).

\bibitem{Castin-Dum} 
Y. Castin and R. Dum, Phys. Rev. Lett.  {\bf 77} 5315-5319 (1996).

\bibitem{Kagan} Yu. Kagan, E. L. Surkov, and G. V. Shlyapnikov,
Phys. Rev. Lett. {\bf 79}, 2604-2607 (1997).

\bibitem{vu_a_ENS} 
P. Rosenbusch, V. Bretin, and J. Dalibard, cond-mat/0206511.

\bibitem{pas_uni} When the condensate is in a hard wall container, 
like typical configurations for superfluid helium, the density 
can be taken to be uniform, except close to the vortex line. 
The velocity field $\vec{v}$ of the condensate can then be determined from the 
continuity equation div $\vec{v}=0$ and the value of $\vec{\mbox{curl}}\,\vec{v}$ which depends
on the vortex line shape only. 
In our case the continuity equation for $\Omega=0$
is div $n\vec{v}=0$ where $n$ is the non-uniform
density, which makes the determination of $\vec{v}$  more difficult.

\bibitem{personal_estimate} The rotation frequency destabilizing the
surface modes of the vortex free condensate scales as $\omega_\perp/\sqrt{l_c}
\sim \omega_\perp (\mu/\hbar\omega_\perp)^{-1/3}$, where the angular momentum
$l_c$ is the cross-over from the collective to the single particle behavior
of the surface modes, see 
F. Dalfovo, S. Giorgini, M. Guilleumas, L. Pitaevskii, and S. Stringari,
Phys. Rev. A {\bf 56}, 3840-3845 (1997).

\bibitem{proviso} One has to perform this limit in a way preserving the
Thomas-Fermi regime. For example setting $\omega_z$ to zero while
keeping $\omega_\perp$ and $Na$ constant is not appropriate. One
can on the contrary keep $\mu/\hbar\omega_\perp$ constant by adjusting $Na$.

\bibitem{general} The same conclusion is obtained in the limit
$\omega_z/\omega_\perp\to 0$ for a linear piecewise ansatz
with more parameters, for example an adjustable shift $x_0$ along $x$ of
the vertical segment or an adjustable angle for the lines going to infinity.

\bibitem{U1} Y. Castin and R. Dum, Phys. Rev. A {\bf 57}, 3008-3021 (1998).
C. Gardiner, Phys. Rev. A {\bf 56}, 1414-1423 (1997).

\bibitem{Girardeau} M.D. Girardeau and R. Arnowitt, Phys. Rev.
{\bf 113}, 755 (1959).

\bibitem{Wigner}
Alice Sinatra, Carlos Lobo and Yvan
Castin, Phys. Rev. Lett. {\bf 87}, 210404 (2001).

\bibitem{gardiner} C. W. Gardiner, {\it Quantum Noise}, Springer-Verlag (1991).

\bibitem{phase_globale}
In principle one should add an overall phase factor
$e^{i\theta}$ to the expression for $\psi$, corresponding to the
eigenvalue of the operator $\hat{A}_\phi$ and
expressing the fact
that the phase $\theta$ of the condensate is fluctuating randomly from
one shot to the other, but this has no physical consequence on
observables conserving the number of particles. A similar phase
factor appears in the Wigner point of view, see \cite{Wigner}.


\bibitem{cartago} Alice Sinatra, Carlos Lobo and Yvan Castin,
J. Phys. B {\bf 35}, 3599-3631 (2002).

\bibitem{verif_curv} 
For a trap aspect ratio $\omega_\perp/\omega_z=15$ and a chemical potential
of the order of $8\hbar\omega_\perp$, we find a maximal curvature of the vortex line
of the order of $2/R_\perp$, where $R_\perp$ is the Thomas-Fermi radius of the
condensate in the $xy$ plane. The curvature term $k^2/4$ inside the logarithm
of the energy functional of \cite{AftaRiv} is therefore at least two times smaller than 
the other, curvature independent term $\rho_{\rm TF}^{1/2}\Delta_\perp
\rho_{\rm TF}^{-1/2}$.

\bibitem{convenience} This definition of $\xi$ is $\sqrt{2}$ times larger than the
usual healing length but turns out to be more convenient in terms of the elimination
of residual factors of 2.

\bibitem{en_principe} In principle and for a given profile
$n$, $\vec{v}$ is determined in a unique way from
the requirement that a single vortex is present, (\ref{eq:rot_connu}),
 and from the continuity equation
(\ref{eq:sta1}), as shown in an appendix of \cite{castindum}. The solution
of (\ref{eq:sta1}) is actually not trivial to find analytically.

\bibitem{different} 
This slightly differs from \cite{AftaRiv}, as we put the full density $n$
rather than the Thomas-Fermi density $n_{\rm TF}$ in the right hand side
of the equation. Since $n$ vanishes on the vortex line, $\vec{B}$
is well defined everywhere.

\bibitem{pourquoi} A physical
interpretation of this fact is that,
$n_{\rm TF}\vec{\Omega}\land\vec{r}$, being the probability current $\vec{j}$
of a cylindrically symmetric solid body of density $n_{\rm TF}$ rotating
at angular velocity $\Omega$ around its symmetry axis $z$, satisfies $\mbox{div}(\vec{j}\,)=0$
because of the continuity equation.

\bibitem{Jackson} J. D. Jackson, {\it Classical Electrodynamics, third edition},
John Wiley (New York, 1999).

\bibitem{terme_de_bord} 
Even without knowing $\vec{v}$ precisely we deduce from the Biot and Savart
formula that the flux of $\vec{v}\land\vec{B}$ indeed vanishes at infinity,
see the appendix \ref{appen:loin}.

\bibitem{plus_facile}
Such an approximation, replacing the function $f$ by unity, if performed in $E_{\rm phase}^{\rm kin}$,
would have resulted in a divergent integral, since $v^2$ diverges as $1/\rho^2$ close to the
vortex line.

\end{thebibliography}
\end{document}